\documentclass[reprint,aps,prx,superscriptaddress,longbibliography]{revtex4-2}
\usepackage[T1]{fontenc}
\usepackage[utf8]{inputenc}
\usepackage[english]{babel}
\usepackage{amsmath}
\usepackage{amssymb}
\usepackage[export]{adjustbox}
\usepackage{bm}
\usepackage{braket}
\usepackage{xcolor}
\usepackage{ytableau}

\usepackage[colorlinks=true,urlcolor=blue,citecolor=blue,linkcolor=blue]{hyperref}

\DeclareMathOperator{\id}{id}
\DeclareMathOperator{\Tr}{Tr}
\DeclareMathOperator{\sign}{sign}
\DeclareMathOperator{\perm}{perm}
\DeclareMathOperator{\imm}{imm}
\DeclareMathOperator{\stab}{stab}
\renewcommand{\i}{\mathrm{i}}
\newcommand{\e}{\mathrm{e}}

\newcommand{\freiburg}{Physikalisches Institut, Albert-Ludwigs-Universit\"at Freiburg,\\ Hermann-Herder-Stra\ss e 3, D-79104 Freiburg, Germany}
\newcommand{\eucor}{EUCOR Centre for Quantum Science and Quantum Computing, Albert-Ludwigs-Universit\"at Freiburg,\\ Hermann-Herder-Stra\ss e 3, D-79104 Freiburg,  Germany}

\begin{document}

\title{Fourier analysis of many-body transition amplitudes and states}
	
	\author{Gabriel Dufour}
	\affiliation{\freiburg}
	\affiliation{\eucor}

	\author{Andreas Buchleitner}
	\affiliation{\freiburg}
	\affiliation{\eucor}

\begin{abstract}
We decompose the counting statistics of many-body interference experiments into contributions associated with distinct irreducible exchange symmetries.
To do so, we perform a Fourier transform over the symmetric group $S_N$ on the collection of $N!$ many-body transition amplitudes connecting two states of a system of $N$ particles.
We apply our formalism to the interference of partially distinguishable bosons and fermions and describe mechanisms responsible for completely destructive interference in many-body systems obeying specific exchange symmetries, including, but not limited to, bosons and fermions.
\end{abstract}

\maketitle


\section{Introduction}

Harmonic analysis is one of the most powerful and versatile tools at the disposal of mathematicians, engineers and physicists alike.
The various flavours of Fourier transforms can notably be put to  use to obtain a compact and interpretable representation of complex signals. In particular, the Parseval-Plancherel identity allows to break down the total energy of a wave into contributions from each frequency component.
More fundamentally, the Fourier transform provides an analysis of the scales on which the function varies by decomposing it into a basis of common eigenfunctions of all translation operators. Differential and difference equations are thereby mapped onto simpler, algebraic ones. 
The Fourier transform also holds a prominent place in quantum mechanics, since it relates the representations of quantum states in position and momentum space, the two being constrained by the Heisenberg uncertainty relation.
In its discrete version, the Fourier transform plays a central role in quantum computing, 
as it is instrumental in providing an exponential speed-up in resolving so-called hidden subgroup problems, including for example Deutsch's problem, Simon's problem and the integer factorization problem \cite{kitaev_quantum_1995,jozsa_quantum_2001}.

All of the above applications can be formulated in the language of representation theory of Abelian (i.e., commutative) groups.
The Abelian group in question is the group of translations of a given set. Each translation induces a linear map on the vector space of functions defined on that set (or, equivalently, of functions defined on the group of translations itself). Since they commute with one another, these maps can be simultaneously diagonalized and the Fourier transform essentially gives the decomposition of a function in the simultaneous eigenbasis.
However, the Fourier transform can also be extended to functions defined on a \textit{non-commutative} finite group. In this case, the action of the group can no longer be simultaneously diagonalized, but representation theory teaches us that it can be broken down into a finite number of elementary operations: the group's irreducible representations.
The Fourier transform decomposes a function over a group in a basis where the group acts irreducibly.
The Fourier transform over non-Abelian finite groups has found applications in probability theory and statistics (e.g., to analyse random processes such as the shuffling of a deck of cards) \cite{diaconis_applications_1988,rockmore_applications_1997}, in the study of ranking schemes (e.g., elections)  \cite{diaconis_generalization_1989,daugherty_voting_2007}, in graph and number theory \cite{terras_fourier_1999}, 
as well as in machine learning tasks displaying symmetries \cite{kondor_group_2008}.
 The  success of the ordinary Fourier transform in practical applications certainly owes a lot to the existence of the Fast Fourier Transform algorithm \cite{cooley_algorithm_1965}.
By extending its principle to the non-Abelian case \cite{clausen_fast_1989,diaconis_efficient_1990},
Fourier transforms over finite groups can be  computed explicitly, despite the rapidly growing size of the involved groups.
  

In this work, we apply the Fourier transform over the symmetric group $S_N$ to the analysis of many-body interference processes in the dynamics of $N$ identical quantum particles.
Many-body interference arises from the inability, in a transition
between two many-body states, to unambiguously identify each particle in the final state with a particle in the initial state. As a consequence, one has to consider all $N!$ ways of connecting the particles in the initial state with those in the final state, each corresponding to a different many-body path taken by the system, each with its own complex transition amplitude.
The probability of observing the transition results from the interference of these $N!$ amplitudes.
Furthermore, more often than not, the interfering particles carry internal degrees of freedom which
don't participate in the dynamics but make the particles partially distinguishable from one another \cite{tichy_fourphoton_2011,ra_nonmonotonic_2013,shchesnovich_partial_2015,tichy_sampling_2015,dittel_waveparticle_2021}. The many-body transition probabilities then additionally depend on these internal states.
On the one hand, uncontrolled partial distinguishability is a source of decoherence in protocols based on many-body interference, calling for dedicated diagnostics and certification tools \cite{walschaers_statistical_2016,giordani_experimental_2018,stanisic_discriminating_2018,giordani_experimental_2020,brunner_manybody_2022,young_atomic_2024,geller_measuring_2025}.
On the other hand, control over the particles' internal states provides a handle to shape their many-body interference patterns \cite{tan_su_2013,deguise_coincidence_2014,tillmann_generalized_2015,menssen_distinguishability_2017,jones_multiparticle_2020,seron_boson_2023,kumar_exchange_2025}.

We will show that the Fourier transform over the symmetric group provides a natural framework to approach questions relating to  many-body interference. In particular, we will use it to express many-body transition probabilities, to systematically assess the impact of partial distinguishability on the many-body dynamics, and to investigate cases of completely destructive interference generalizing the Hong-Ou-Mandel effect \cite{hong_measurement_1987,bouchard_twophoton_2021}.
Since Weyl and  Wigner \cite{weyl_theory_1950,wigner_group_1959}, group theoretical methods have had a long tradition in quantum mechanics---including, but not limited to the description of quantum statistics.
In the study of many-body interference phenomena, previous works have made use of Schur-Weyl duality, which relates irreducible representations of the symmetric group to those of the unitary group
\cite{adamson_multiparticle_2007,adamson_detecting_2008,rowe_dual_2012,tan_su_2013,deguise_coincidence_2014,tillmann_generalized_2015,moylett_quantum_2018,stanisic_discriminating_2018,khalid_permutational_2018,dufour_manybody_2020,spivak_generalized_2022,brunner_manybody_2023,geller_characterization_2024}.
The originality of the present approach is that it makes no reference to the representations of the unitary group. The advantage is that we only need to work with the finite group $S_N$, whose irreducible representations can be computed once and for all since they do not depend on the dimension of the single-particle Hilbert space. In practice, we use Young's orthogonal representation as defined in \cite{kondor_group_2008} and implemented in the SnPy  package  \footnote{H. Pan, SnPy  package, 
	\url{github.com/horacepan/snpy} (downloaded in June 2023)}.

We start by examining transition amplitudes between symmetrized many-body states and show that these involve convolution products and inner products of functions defined on the symmetric group, motivating the introduction of a Fourier transform over $S_N$.
In Sec.~\ref{sec:FT}, we give the definition and main properties of the Fourier transform over a finite group, and introduce some useful tools from group theory. 
We then proceed with the Fourier analysis of the set of transition amplitudes connecting two $N$-particle states in Sec.~\ref{sec:a}. These transition amplitudes are combined into Fourier components associated with specific particle-exchange symmetries, including the familiar bosonic and fermionic symmetries, but also more exotic ``mixed'' symmetries \cite{messiah_symmetrization_1964,tichy_extending_2017,kumar_exchange_2025}.
In Sec.~\ref{sec:countstat}, we study the counting statistics of particles obeying exchange symmetries generalizing those of indistinguishable bosons and fermions. In particular, the transition probabilities of partially distinguishable particles are expressed in terms of the Fourier transforms of the transition amplitudes, on the one hand, and of a function encoding partial distinguishability, on the other hand.
As an application of our formalism, in Sec.~\ref{sec:suppr}, we state conditions for the contribution of a given symmetry type to a transition probability to vanish. This generalizes the suppression laws  governing completely destructive interference of bosons and fermions \cite{lim_generalized_2005,tichy_zerotransmission_2010,tichy_manyparticle_2012,tichy_stringent_2014,crespi_suppression_2015,crespi_suppression_2016,dittel_manybody_2017,dittel_totally_2018,dittel_totally_2018a} to other particle-exchange symmetries. 
We illustrate these results by performing the harmonic analysis of transition amplitudes in the Fourier interferometer (i.e., for an evolution given by the discrete Fourier transform unitary), where we observe a wealth of such suppressed transitions.

\section{From many-body interference to Fourier analysis}
\label{sec:MBItoFourier}

Before delving into the formalism of the Fourier transform over finite groups, we give a hint of  how symmetry considerations enter the study of many-body-interference phenomena and motivate why Fourier analysis is a suitable tool to approach this problem.
We consider a system of $N$ identical particles, each endowed with an $M$-dimensional single-particle Hilbert space  $\mathcal{H}$ spanned by a set of orthogonal modes $\{\ket{0},\dots \ket{M-1}\}$.
We adopt a first-quantization notation, in which the $N$-particle Hilbert space  $\mathcal{H}^{\otimes N}$, of dimension $M^N$, has basis vectors 
\begin{align}\label{basisvec}
	\ket{\bm{m}}=\ket{m_1}\otimes\ket{m_2}\otimes \dots \ket{m_N},
\end{align}
 with $m_\alpha\in \{0,\dots M-1\}$ for $\alpha=1,\dots N$ (note that we count the modes from $0$ to $M-1$ while particles are labelled from $1$ to $N$). 

The many-body system is assumed to undergo a non-interacting, unitary evolution, as realized, e.g., by a linear $M$-port interferometer. The evolution thus maps a basis state $\ket{\bm{m}}$ to 
\begin{align}\label{reprU}
	U^{\otimes N} \ket{\bm{m}} =(U\ket{m_1})\otimes(U\ket{m_2})\otimes \dots (U\ket{m_N}),	  
\end{align}
where $U\in U(M)$ is a unitary transformation of $\mathcal{H}$.
The many-body transition amplitude between an input basis state $\ket{\bm{i}}$ and an output basis state $\ket{\bm{o}}$ thus factorizes as
\begin{align}
	\braket{\bm{o}|U^{\otimes N}|\bm{i}} = \prod_{\alpha=1}^N \braket{o_\alpha|U|i_\alpha}.
\end{align}

Although transition amplitudes are easily evaluated in the above product basis, there are good reasons to prefer working in another, \emph{symmetrized} (in a sense that we will make more precise) basis. 
The first reason is mathematical: we can take advantage of the fact that \textit{the same} unitary evolution operator $U$ is being applied to each particle, i.e., the many-body evolution is invariant under particle permutations. When expressed in the appropriate basis, the $N$-particle evolution operator $U^{\otimes N}$ thus takes a block-diagonal form, with  blocks corresponding to uncoupled symmetry sectors.
The second---and not completely unrelated---reason is physical: the symmetrization postulate of quantum mechanics posits that physical states of identical particles must either be completely symmetric under exchange of particles---in the case of bosons---or completely antisymmetric---in the case of fermions \cite{messiah_symmetrization_1964}. At first sight, this means that only two  of the aforementioned symmetry sectors are relevant: those corresponding to bosonic or fermionic states. However, as we will argue later on, states with a more involved symmetry are also of interest.

To illustrate these points, we start with a well-known example: the Hong-Ou-Mandel interference of two particles at a beam splitter \cite{hong_measurement_1987,bouchard_twophoton_2021}. We consider a two-particle, two-mode system ($N=M=2$) and the beam splitter unitary 
\begin{align}\label{Ubs}
	U=\frac{1}{\sqrt{2}}\begin{pmatrix}
		1 & 1\\ 1 & -1
	\end{pmatrix}.
\end{align}
In the product basis $\{\ket{0,0},\,\ket{0,1},\,\ket{1,0},\,\ket{1,1} \}$ of $\mathcal{H}^{\otimes 2}$, the matrix representation of the two-body evolution operator $U\otimes U$ reads
\begin{align}
	\frac12\begin{pmatrix}
		1 & 1 & 1 &1\\
		1 & -1 & 1 & -1\\
		1 & 1  & -1 & -1\\
		1 & -1 & -1  & 1
	\end{pmatrix}.
\end{align}
However, if we represent the same operator in the following basis: 
\begin{align}
	\Big\{\ket{0,0},\, \ket{\Psi^{(+)}}&= \frac{1}{\sqrt{2}}\left( \ket{0,1} + \ket{1,0} \right),\, \ket{1,1},\notag \\
\ket{\Psi^{(-)}}&=\frac{1}{\sqrt{2}}\left( \ket{0,1} - \ket{1,0} \right)\Big\},
\end{align}
the first three states of which  are  symmetric under particle exchange, and the last one antisymmetric, we instead obtain
\begin{align}
	\frac12\begin{pmatrix}
			1 & \sqrt{2} & 1 & 0 \\
			\sqrt{2} & 0 & -\sqrt{2} & 0 \\
			1 & -\sqrt{2} & 1 & 0 \\
			0 & 0 & 0 & -2 
	\end{pmatrix}.
\end{align}
In this symmetrized basis, the evolution operator displays more structure. First of all, there is no coupling between states with different symmetries, as  evidenced by the zeroes in the last row and column of the above matrix. Moreover, in the symmetric sector, we additionally observe a vanishing transition amplitude $\braket{\Psi^{(+)}|U\otimes U|\Psi^{(+)}}=0$. This is of course the famous suppression of coincidence events (where one particle exits in each mode) when two bosons enter in distinct modes of the beam splitter. This instance of totally destructive many-body interference is a consequence of the \emph{joint} symmetry of the state and of the interferometer unitary $U$ \cite{dittel_totally_2018,dittel_totally_2018a}.
While this is certainly a striking effect, the two-particle case is in a sense trivial because the bosonic and fermionic states span the entire Hilbert space. This is no longer true in systems of $N>2$ particles, where other exchange symmetries appear.

Let us consider for example $N=3$ particles and $M=3$ modes and restrict ourselves to states with one particle per mode. A symmetrized basis of this six-dimensional subspace is given by
\begin{widetext}
\begin{align}\label{3psymstates}
	\Big\{\ket{\Psi^{(+)}}&=\frac{1}{\sqrt{6}}\left(\ket{0, 1, 2}+ \ket{0, 2, 1}+ \ket{1, 0, 2}+ \ket{2, 0, 1}+ \ket{1, 2, 0}+ \ket{2, 1, 0} \right),\notag \\
	\ket{\Psi^{(0)}_{1}}&=\frac{1}{\sqrt{6}}\left(2\ket{0, 1, 2}- \ket{2, 0, 1}- \ket{1, 2, 0} \right), \ket{\Psi^{(0)}_{2}}=\frac{1}{\sqrt{2}}\left( \ket{1, 0, 2}-\ket{2, 1, 0} \right),\notag \\
	\ket{\Psi^{(0)}_{3}}&=\frac{1}{\sqrt{6}}\left(2 \ket{0, 2, 1}-\ket{1, 0, 2}- \ket{2, 1, 0} \right),
	\ket{\Psi^{(0)}_{4}}=\frac{1}{\sqrt{2}}\left( \ket{2, 0, 1}-\ket{1, 2, 0} \right),\notag \\
	\ket{\Psi^{(-)}}&=\frac{1}{\sqrt{6}}\left(\ket{0, 1, 2}- \ket{0, 2, 1}- \ket{1, 0, 2}+ \ket{2, 0, 1}+ \ket{1, 2, 0}- \ket{2, 1, 0} \right)\Big\}.
\end{align} 
\end{widetext}
The first and last states are completely symmetric and antisymmetric, respectively, while the four remaining states have a more elaborate symmetry.
We now consider the evolution generated by the following \emph{tritter} unitary
\begin{align}
	U=\frac{1}{\sqrt{3}}\begin{pmatrix}
		1 & 1 & 1\\ 1 & j & j^2\\ 1 & j^2 & j 
	\end{pmatrix}\qquad \text{with}\quad j = \exp(2\i\pi/3),
\end{align}
which generalizes the beam splitter unitary \eqref{Ubs} to three modes.
The only non-vanishing matrix elements of $U^{\otimes 3}$ in the above basis are found to be
\begin{align}
	\braket{\Psi^{(+)}|U^{\otimes 3}|\Psi^{(+)}}&=-\frac{1}{\sqrt{3}}\\ \intertext{and}
	\braket{\Psi^{(-)}|U^{\otimes 3}|\Psi^{(-)}}&=-\i. 
\end{align}
Not only do transition amplitudes between states with different symmetries vanish, as expected, but all transitions between states which are neither bosonic nor fermionic [labelled $(0)$ above] are also suppressed.
Like the Hong-Ou-Mandel effect, this is an instance of completely destructive many-body interference, arising from the joint symmetry of the tritter unitary and of the many-body state. However, in this case, the state's symmetry is neither bosonic nor fermionic.

Let us now take a step back and consider transition amplitudes between two $N$-particle states which have been subjected to some form of symmetrization (again, we will make this more precise). To this end, we build symmetrization operators which map a product state to a superposition of states obtained from it by permutation of its factors.
We can generically write such an operator as follows:
\begin{align}\label{eq:hatc}
	\hat{c}(R)=\sum_{\sigma\in S_N} c(\sigma) \hat{R}(\sigma).
\end{align}
The reason for denoting this operator $\hat{c}(R)$ will become clear in the following, so let us start by explaining the right-hand side of this definition: the sum runs over permutations $\sigma$ in  the symmetric group $S_N$, the $c(\sigma)$ are complex coefficients and the operators $\hat{R}(\sigma)$ act on the $N$-particle basis states \eqref{basisvec} by reordering the factors of the tensor product according to the permutation $\sigma$:
\begin{align}\label{eq:reprSN}
	\hat{R}(\sigma) \ket{\bm{m}}&=\hat{R}(\sigma)\left(  \ket{m_1}\otimes\ket{m_2}\otimes \dots \ket{m_N} \right) \\ &=\ket{m_{\sigma^{-1}(1)}}\otimes\ket{m_{\sigma^{-1}(2)}}\otimes \dots \ket{m_{\sigma^{-1}(N)}}.\notag
\end{align}
The map $R:\sigma\mapsto \hat{R}(\sigma)$ thus defines a unitary representation of the symmetric group, meaning that
for the composition $\sigma\circ\tau$ of two  permutations,  we have $\hat{R}(\sigma\circ\tau) =\hat{R}(\sigma) \hat{R}(\tau)$, 
while for the inverse $\sigma^{-1}$ of a permutation,  $\hat{R}(\sigma^{-1})=\hat{R}(\sigma)^{-1}=\hat{R}(\sigma)^\dagger$ holds.
The usual symmetrization and antisymmetrization operators, i.e., the projectors on bosonic and fermionic states, are of course included as special cases: the former corresponds to the choice $c(\sigma)=1/N!$, while for the latter we take $c(\sigma)=\sign(\sigma)/N!$, with $\sign(\sigma)$ the signature of the permutation $\sigma$. However,  as we have seen in the above example, more general forms of symmetrization can also be relevant for more than two particles.

Given two product basis states $\ket{\bm{i}}$ and $\ket{\bm{o}}$ and two symmetrization operators $\hat{c}(R)$ and $\hat{d}(R)$ [defined as in Eq.~\eqref{eq:hatc}], 
the transition amplitude between $\hat{c}(R)\ket{\bm{i}}$ and $\hat{d}(R)\ket{\bm{o}}$ can be written as
\begin{align}\label{transampcd}
	&\braket{\bm{o}|\hat{d}(R)^\dagger U^{\otimes N} \hat{c}(R)|\bm{i}}\notag\\
	&\qquad =\sum_{\sigma,\tau\in S_N} d(\sigma)^* c(\tau)  \braket{\bm{o}|\hat{R}(\sigma)^\dagger U^{\otimes N} \hat{R}(\tau)|\bm{i}}.
\end{align}
Since the permutation operators $\hat{R}(\tau)$ commute with the evolution operator $U^{\otimes N}$, and using the fact that $R$ is a unitary representation, we can rewrite 
\begin{align}\label{firstdefa}
 \braket{\bm{o}|\hat{R}(\sigma)^\dagger U^{\otimes N} \hat{R}(\tau)|\bm{i}}&= \braket{\bm{o}|\hat{R}(\tau^{-1}\circ\sigma)^\dagger U^{\otimes N} |\bm{i}}\notag\\
 &= a(\tau^{-1}\circ\sigma),
\end{align}
which defines the \emph{transition amplitude function} $a$ associated with the unitary $U$ and the pair of states $\ket{\bm{i}}$ and $\ket{\bm{o}}$ (these dependences are left implicit to avoid overburdening the notation). 
In terms of the three functions $a$, $c$ and $d$---all three defined over $S_N$---the transition amplitude thus reads
\begin{align}\label{convol+inner}
	&\braket{\bm{o}|\hat{d}(R)^\dagger U^{\otimes N} \hat{c}(R)|\bm{i}}\notag\\
	&\qquad
=\sum_{\sigma\in S_N}d(\sigma)^* \left( \sum_{\tau\in S_N}  c(\tau)  a(\tau^{-1}\circ\sigma) \right).  
\end{align}
The term in parentheses has the form of a \textit{convolution product} between $c$ and $a$.
With the sum over $\sigma$, we then take the \textit{inner product} of  the resulting function with $d$.
The appearance of these products calls for the application of a Fourier transform. Indeed, one defining property of such a transformation is that it maps convolution products to ordinary products. Moreover, by the Parseval-Plancherel identity, it also preserves the inner product. In the following, we introduce the appropriate Fourier transform for functions defined over a finite group, such as $S_N$, and elaborate on its effect on convolution and inner products.

\section{Fourier transform over a finite group}
\label{sec:FT}

The ordinary Fourier transform is the expansion of a function in the basis of common eigenfunctions of shift operators, i.e., operators which shift a function's argument. Such a basis exists because shift operators commute with one another. If we instead consider a non-Abelian group of transformations of the function's argument, this is no longer true. However, we can find a basis in which the group's action takes the simplest possible form. The Fourier transform over a finite group performs the corresponding basis change. 
In this section, we first give the definition of the Fourier transform over a finite group and state its main properties. We then introduce some of the useful mathematical objects which occur in the rest of this work. We assume that the reader is familiar with the basic concepts of group theory and refer to \cite{hamermesh_group_1989,chen,sengupta} for a more detailed exposition of the representation theory of finite groups.

\subsection{Definitions}
\label{sec:def}

Consider a group $G$ with a finite number $|G|$ of elements, whose group operation we denote $\circ$. 
A $d$-dimensional matrix representation $\rho$ of $G$ is a map
\begin{align}\label{eq:representation}
	\rho: \sigma\mapsto \hat{\rho}(\sigma)
\end{align} 
which associates to every group element $\sigma\in G$ a $d\times d$ invertible complex matrix $\hat{\rho}(\sigma)$ and which is compatible with the group operation: for all pairs of elements $\sigma,\tau\in G$, the matrix associated with $\sigma\circ\tau$ is obtained by matrix multiplication of the matrices associated with $\sigma$ and $\tau$:
\begin{align}\label{morphism}
 \hat{\rho}(\sigma\circ\tau)=\hat{\rho}(\sigma)\hat{\rho}(\tau).
\end{align}
In this work, we will only be considering \textit{unitary} representations, such that the matrix associated with the inverse $\sigma^{-1}$ of a group element $\sigma\in G$ is the Hermitian conjugate of $\hat{\rho}(\sigma)$:
\begin{align}
	\hat{\rho}(\sigma^{-1})=\hat{\rho}(\sigma)^{-1}=\hat{\rho}(\sigma)^\dagger.
\end{align}

Any given representation $\rho$ of a finite group $G$ can be broken down into a finite set of irreducible unitary representations (irreps) $\rho^{(\lambda)}$, of dimension $d_\lambda$.
This means that, in an appropriately chosen basis, the matrices $\hat{\rho}(\sigma)$ all share the same block-diagonal structure
\begin{align}\label{irrepdecomp}
	\hat{\rho}(\sigma)\simeq \bigoplus_\lambda \mathbb{I}_{m_\lambda}\otimes \hat{\rho}^{(\lambda)}(\sigma),
\end{align}
with the  $d_\lambda\times d_\lambda$ block  $\hat{\rho}^{(\lambda)}(\sigma)$ appearing $m_\lambda$ times (here, $\simeq$ denotes equality up to a basis change and $\mathbb{I}_{m}$ is the $m\times m$ identity matrix).
The unitary irrep matrices $\hat{\rho}^{(\lambda)}(\sigma)$ cannot be further jointly block-diagonalized  and they are unique up to a change of orthonormal basis. Here and in the following, sums over $\lambda$ are understood to run over all irreps of the group $G$ under consideration.

We now come to the definition of the Fourier transform over a finite group $G$. Let 
\begin{align}
	f:\begin{cases}
		G&\to \mathbb{C}\\
		\sigma&\mapsto f(\sigma)
	\end{cases}
\end{align}
be a complex-valued function over $G$. The Fourier transform $\hat{f}(\rho)$ of $f$ over $G$ at the $d$-dimensional representation $\rho$ is the $d\times d$ matrix 
\begin{align}\label{FT}
	\hat{f}(\rho)=\sum_{\sigma\in G} f(\sigma) \hat{\rho}(\sigma).
\end{align}
For example, the symmetrization operator $\hat{c}(R)$ defined in Eq.~\eqref{eq:hatc} is the Fourier transform of the function $c$ at the representation $R$, which explains our choice of notation.
Note that we use hats to denote representation matrices, such as $\hat{\rho}(\sigma)$ in Eq.~\eqref{morphism}, as well as Fourier transforms, which are linear combinations of such matrices, for example $\hat{f}(\rho)$ in Eq.~\eqref{FT}.
The Fourier transform of a function $f$ at the irrep with label $\lambda$, $\hat{f}(\rho^{(\lambda)})$, will simply be denoted $\hat{f}(\lambda)$ for the sake of conciseness. The knowledge of the Fourier transform $\hat{f}(\lambda)$ of $f$ at all irreps $\lambda$ of $G$ allows to recover the original function through the inverse transform
\begin{align}\label{IFT}
	f(\sigma)=\sum_{\lambda}  \frac{d_\lambda}{|G|} \Tr \left[ \hat{\rho}^{(\lambda)}(\sigma)^\dagger \hat{f}(\lambda)\right].
\end{align}
Here and in the following, sums over $\lambda$ are assumed to run over all inequivalent irreps of $G$.
As we show in greater detail in Appendix \ref{app:IFT}, the individual entries of the irrep matrices $\hat{\rho}^{(\lambda)}(\sigma)$ form an orthogonal basis of the space of complex-valued functions  on $G$, and Eq.~\eqref{IFT} is the expansion of $f$ in that basis.

\begin{figure*}
	\newlength{\imwidth}
	\setlength{\imwidth}{2.4cm}
	\centering
	
	{\large $\bm{\sigma\mapsto \tau\circ\sigma}$ }

	\begin{tabular}{c c c c c c}
		\multicolumn{2}{l}{Cyclic group of order 6}&&&&\\
		\includegraphics[width=\imwidth]{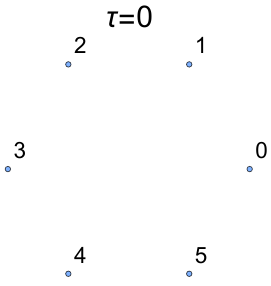}\quad &
		\includegraphics[width=\imwidth]{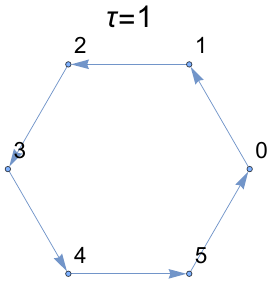}\quad &
		\includegraphics[width=\imwidth]{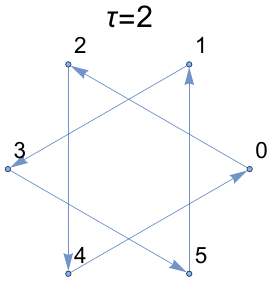}\quad &
		\includegraphics[width=\imwidth]{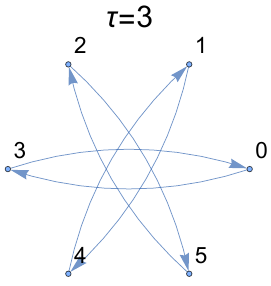}\quad &
		\includegraphics[width=\imwidth]{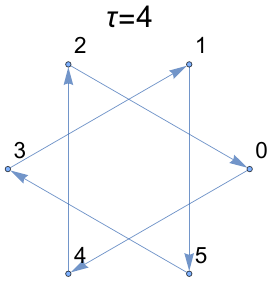}\quad &
		\includegraphics[width=\imwidth]{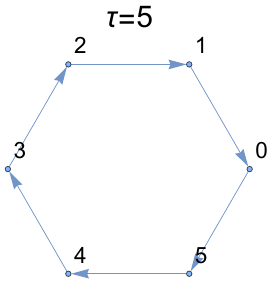}\\
				&&&&&\\	
		\multicolumn{2}{l}{Symmetric group $S_3$} &&&&\\
		\includegraphics[width=\imwidth]{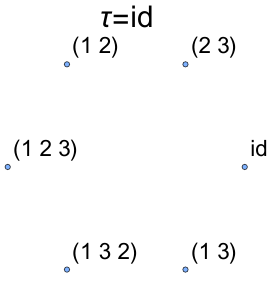}\quad &
		\includegraphics[width=\imwidth]{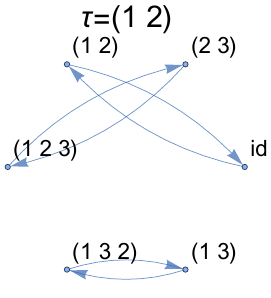}\quad &
		\includegraphics[width=\imwidth]{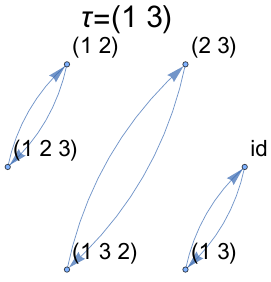}\quad &
		\includegraphics[width=\imwidth]{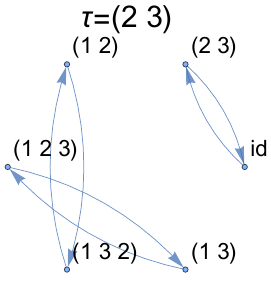}\quad &
		\includegraphics[width=\imwidth]{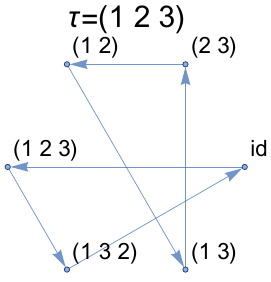}\quad &
		\includegraphics[width=\imwidth]{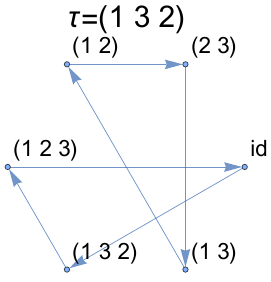}
	\end{tabular}
	\caption{\label{Z6S3} Graphical representations of the multiplication tables of the (Abelian, i.e., commutative) cyclic group of order 6 (top) and (non-Abelian) group $S_3$ (bottom). The group elements $\sigma$ are represented by points on a circle. For each group element $\tau$, the map $\sigma\mapsto \tau\circ\sigma$ is depicted by arrows connecting these points.}
\end{figure*}

We recover the ordinary (discrete) Fourier transform in the case where $G$ is a cyclic group.
Without loss of generality, we can take $G=\{0,1, \dots |G|-1\}$, with the group operation given by addition modulo $|G|$:
\begin{align}\label{addlaw}
	\sigma\circ\tau = \sigma+\tau \mod |G|.
\end{align}
The corresponding multiplication table for $|G|=6$ is schematically represented in the top row of Fig.~\ref{Z6S3}. The group elements $\sigma$ are represented by six points on a circle. We connect these points according to the map 
$\sigma\mapsto \tau\circ\sigma$ for each $\tau\in G$. In that case, compositions by a group element correspond to translations along the ring, which commute with one another. 
Since the commutativity of the group operation \eqref{addlaw} carries over to a representation $\rho$ of $G$, all representation matrices $\hat{\rho}(\sigma)$, for $\sigma\in G$, can be simultaneously diagonalized, such that, by Eq.~\eqref{irrepdecomp}, all irreps of $G$ are one-dimensional .
 Specifically, for the cyclic group, the irreps are given by 
$\hat{\rho}^{(\lambda)}(\sigma)=\exp(2 \i \pi \lambda \sigma/ |G|)$, where both $\sigma$ and $\lambda$ are integers running from $0$ to $|G|-1$.
The Fourier transform [Eq.~\eqref{FT}] and its inverse [Eq.~\eqref{IFT}] then take the familiar forms
\begin{align}\label{DFT}
	&\hat{f}(\lambda)=\sum_{\sigma=0}^{|G|-1} f(\sigma)\exp(2 \i \pi \lambda \sigma/ |G|)\\
	\intertext{and}
	&f(\sigma)= \frac{1}{|G|}\sum_{\lambda=0}^{|G|-1}    \hat{f}(\lambda) \exp(-2 \i \pi \lambda \sigma/ |G|).    
\end{align}

As a simple example of the non-Abelian case, let us consider the group of permutations of three elements, i.e., the symmetric group $S_3$. It has $3!=6$ elements $\sigma$, which we write using cycle notation ($\id$ being the identity permutation). 
We represent the multiplication table of $S_3$ in the bottom row of Fig.~\ref{Z6S3}, where it is apparent that the various transformations do not commute in general. Since it is non-commutative, $S_3$ must therefore possess an irrep $\lambda$ with $d_\lambda>1$. This is the two-dimensional \textit{standard} irrep. The group $S_3$ also has two one-dimensional irreps, the \textit{trivial} and \textit{sign} irreps (irreps of symmetric groups will be discussed in more detail in Sec.~\ref{sec:irreps}). The irrep matrices $\hat{\rho}^{(\lambda)}(\sigma)$ in Young's orthogonal representation are given in Table~\ref{tab:irrepsS3}. 

\begin{table*}
	\begin{tabular}{c | c c c c c c c}
		$\sigma$        &
		 $\id$          &
		  $(1\, 2)$     & 
		  $(1\, 3)$     & 
		  $(2\, 3)$     & 
		  $(1\, 2\, 3)$ & 
		  $(1\, 3\, 2)$ &  \\
		\hline
		$\hat{\rho}^{(\mathrm{triv})}(\sigma) $ & $(1)$ & $(1)$ & $(1)$ & $(1)$ & $(1)$ & $(1)$\\
		$\hat{\rho}^{(\mathrm{stand})}(\sigma) $ &
		 $\ \begin{pmatrix} 1 & 0\\ 0 & 1\end{pmatrix}$& 
		 $\ \begin{pmatrix} -1 & 0\\ 0 & 1\end{pmatrix}$& 
		 $\ \dfrac12\begin{pmatrix} 1 & -\sqrt{3}\\ -\sqrt{3} & -1\end{pmatrix}$&
		 $\ \dfrac12\begin{pmatrix} 1 & \sqrt{3}\\ \sqrt{3} & -1\end{pmatrix}$&
		 $\ \dfrac12\begin{pmatrix} -1 & -\sqrt{3}\\ \sqrt{3} & -1\end{pmatrix}$&
		 $\ \dfrac12\begin{pmatrix} -1 & \sqrt{3}\\ -\sqrt{3} & -1\end{pmatrix}$\\
		$\hat{\rho}^{(\mathrm{sign})}(\sigma) $ &    $(1)$ & $(-1)$ & $(-1)$ & $(-1)$ & $(1)$ & $(1)$
	\end{tabular}
	\caption{\label{tab:irrepsS3} Irreducible representation matrices $\hat{\rho}^{(\lambda)}(\sigma)$ for the three irreps $\lambda$ and six permutations $\sigma$ of $S_3$.}
\end{table*}

\subsection{Basic operations}

The Fourier transform over a finite group $G$ [Eq.~\eqref{FT}] enjoys the following properties:
\begin{itemize}

\item The  \textit{convolution product} $f*g$ of two functions $f$ and $g$, defined [recall the right-hand side of Eq.~\eqref{convol+inner}] by 
\begin{align}\label{convol}
\forall \sigma \in G,\qquad 	(f * g )(\sigma) &= \sum_{\tau\in G} f(\sigma\circ \tau^{-1}) g(\tau) \notag \\
&= \sum_{\tau\in G} f(\tau) g(\tau^{-1}\circ\sigma),
\end{align}
is mapped to the matrix product of their Fourier transforms, i.e., for any representation $\rho$ of $G$, 
\begin{align}\label{convolFT}
	\widehat{f*g}(\rho)= \hat{f}(\rho) \hat{g}(\rho).
\end{align}
Indeed, with the definition \eqref{FT} of the Fourier transform, we have
\begin{align}
	\widehat{f*g}(\rho)=\sum_{\sigma\in G} \left( \sum_{\tau\in G} f(\sigma\circ \tau^{-1})g(\tau)\right) \hat{\rho}(\sigma).
\end{align}
We can now replace the sum over $\sigma$ by a sum over $\pi=\sigma\circ \tau^{-1}$ and  use the factorization $\hat{\rho}(\sigma)=\hat{\rho}(\pi)\hat{\rho}(\tau)$ to obtain
\begin{align}
	\widehat{f*g}(\rho)&=\left( \sum_{\pi\in G} f(\pi) \hat{\rho}(\pi) \right) \left( \sum_{\tau\in G}g(\tau)\hat{\rho}(\tau)\right)\notag\\&= \hat{f}(\rho) \hat{g}(\rho).
\end{align}

\item Let $\delta_\tau$ be the function which evaluates to one at a given $\tau\in G$ and to zero elsewhere, i.e., $\delta_\tau(\sigma)=\delta_{\tau,\sigma}$ for all $\sigma\in G$. Taking its convolution product with a function $f$ applies a \textit{shift} $\tau$ to the argument of $f$. Namely, with Eq.~\eqref{convol},
\begin{align}\label{shift}
\forall \sigma \in G, \qquad	(\delta_\tau * f)(\sigma)= f(\tau^{-1}\circ\sigma).
\end{align}
Applying Eq.~\eqref{convolFT}, we see that the Fourier transform of the shifted function is obtained by multiplying $\hat{f}(\rho)$ by $\hat{\delta}_\tau(\rho)=\hat{\rho}(\tau)$:
\begin{align}\label{actSn}
	\widehat{\delta_\tau * f}(\rho)= \hat{\rho}(\tau) \hat{f}(\rho).
\end{align}

\item The \textit{inner product} $(f , g )$ of two functions $f$ and $g$ is defined [recall, again, the right-hand side of Eq.~\eqref{convol+inner}] by 
\begin{align}\label{scalarprod}
	(f , g ) = \sum_{\sigma\in G} f(\sigma)^* g(\sigma).
\end{align}
To express it in terms of Fourier transforms of $f$ and $g$, we use the inverse Fourier transform formula \eqref{IFT} for $g(\sigma)$ and write
\begin{align}
	&(f,g)=\sum_{\sigma\in G} f(\sigma)^* 
	\sum_{\lambda}  \frac{d_\lambda}{|G|} \Tr \left[ \hat{\rho}^{(\lambda)}(\sigma)^\dagger \hat{g}(\lambda)\right]\notag \\
	&\ =
	\sum_{\lambda}  \frac{d_\lambda}{|G|} \Tr \left[ \left(\sum_{\sigma\in G} f(\sigma)^*  \hat{\rho}^{(\lambda)}(\sigma)^\dagger\right) \hat{g}(\lambda)\right].
\end{align}
Recognizing the Hermitian conjugate of the Fourier transform of $f$ [Eq.~\eqref{FT}] in the round brackets, we arrive at the 
 Parseval-Plancherel identity:
\begin{align}\label{PPid}
(f , g ) = \sum_\lambda \frac{d_\lambda}{|G|} \Tr \left[ \hat{f}(\lambda)^\dagger \hat{g}(\lambda)\right].
\end{align}
The inner product of two functions is thereby expressed as the (weighted) sum of  Hilbert-Schmidt products of their Fourier transforms at each irrep.
In particular, 
\begin{align}
\label{spectdens}
\Tr \left[ \hat{f}(\lambda)^\dagger \hat{f}(\lambda)\right]
\end{align}
 can be interpreted as  the \textit{spectral density} of $f$ at irrep $\lambda$.

\end{itemize}

As already hinted at in Sec.~\ref{sec:MBItoFourier}, the convolution product \eqref{convol} and the inner product \eqref{scalarprod} appear repeatedly in calculations of physical quantities describing systems of $N$ identical particles, with $G=S_N$. The identities \eqref{convolFT} and \eqref{PPid} will therefore be instrumental to express these quantities in a concise and symmetry-resolved manner.

\subsection{Class functions and projectors on irreps}
\label{sec:class}
Class functions are defined over the conjugacy classes of $G$, i.e., they take the same value at $\sigma\in G$ and at all elements $\sigma'=\tau\circ\sigma\circ\tau^{-1}$, for $\tau \in G$, of its conjugacy class. 
We can also express this condition as follows: for $c$ a class function, it holds that
\begin{align}
	 c(\tau\circ\sigma)=c(\sigma\circ\tau) \qquad \forall \sigma,\tau \in G.
\end{align}
Using this relation in the definition of the convolution product Eq.~\eqref{convol}, a class function $c$ is found to commute with any function $f$ on $G$, in the sense that 
\begin{align}
	c*f =f*c.
\end{align} 
As a consequence of Schur's lemma \cite{hamermesh_group_1989,chen,sengupta}, the Fourier transform $\hat{c}(\lambda)$ of a class function $c$ at an irrep $\lambda$ is proportional to the identity, i.e.
\begin{align}\label{classfunc}
	\hat{c}(\lambda)= c_\lambda \hat{\mathbb{I}}_{d_\lambda}
\end{align}
for a constant $c_\lambda$.
A particularly interesting case is when $c_\lambda$ vanishes for all but one irrep, which we call $\mu$. Suppose for example that we have a class function $P^{(\mu)}$ such that 
\begin{align}\label{hatPmu}
		\hat{P}^{(\mu)}(\lambda)= \delta_{\mu,\lambda}\hat{\mathbb{I}}_{d_\lambda}.
\end{align}
Applying the inverse Fourier transform \eqref{IFT}, we find that
\begin{align}\label{hatPmuchimu}
	P^{(\mu)}(\sigma)=   \frac{d_\mu}{|G|} \Tr \left[ \hat{\rho}^{(\mu)}(\sigma)^\dagger \right]= \frac{d_\mu}{|G|}\chi^{(\mu)}(\sigma)^*,
\end{align}
where we recognize the irreducible character 
\begin{align}
	\chi^{(\mu)}(\sigma)= \Tr \left[ \hat{\rho}^{(\mu)}(\sigma) \right].
\end{align}
Up to multiplication by a constant and complex conjugation, $P^{(\mu)}$ is just the character $\chi^{(\mu)}$, which is indeed a class function.

If we now take the Fourier transform of $P^{(\mu)}$ at a \textit{reducible} representation $\rho$, we obtain the projector on all irreducible subrepresentations of type $\mu$ of the representation $\rho$:
\begin{align}\label{isotyp}
	\hat{P}^{(\mu)}(\rho)=  \frac{d_\mu}{|G|}\sum_{\sigma\in G} \chi^{(\mu)}(\sigma)^*\hat{\rho}(\sigma).
\end{align}
Such projectors obey
\begin{align}\label{propproj}
	&	\hat{P}^{(\mu)}(\rho)^\dagger=\hat{P}^{(\mu)}(\rho), \qquad 
 \hat{P}^{(\mu)}(\rho) \hat{P}^{(\lambda)}(\rho)= \delta_{\mu,\lambda} \hat{P}^{(\mu)}(\rho) \notag\\
 &   \quad \text{and} \quad \sum_\mu \hat{P}^{(\mu)}(\rho)=\hat{\mathbb{I}}_d,
\end{align}
i.e., they decompose the full representation space into mutually orthogonal subspaces.
In the basis where the matrices $\hat{\rho}(\sigma)$ take the block-diagonal form \eqref{irrepdecomp}, we have 
\begin{align}
	\hat{P}^{(\mu)}(\rho)\simeq \bigoplus_\lambda \delta_{\lambda,\mu} \hat{\mathbb{I}}_{m_\lambda}\otimes\hat{\mathbb{I}}_{d_\lambda}.
\end{align}
Taking the trace of the above relation, we recover the formula for the multiplicity $m_\mu$ of a sub-irrep $\mu$ of the representation $\rho$ in terms of this representation's character,
\begin{align}\label{character}
	\chi(\sigma)=\Tr(\hat{\rho}(\sigma)),
\end{align}
namely, 
\begin{align}\label{mult}
	m_\mu =\frac{1}{|G|}\sum_{\sigma\in G} \chi^{(\mu)}(\sigma)^* \chi(\sigma).
\end{align}
In particular, every group $G$ possesses a one-dimensional trivial irrep $\rho^{(\text{triv})}$ such that  $\hat{\rho}^{(\text{triv})}(\sigma)=\chi^{(\text{triv})}(\sigma)=1$  for all $\sigma\in G$. With Eq.~\eqref{isotyp}, the projector on the trivial components of a representation $\rho$ is given by
\begin{align}\label{projtriv}
	\hat{P}^{(\text{triv})}(\rho)= \frac{1}{|G|}\sum_{\sigma\in G} \hat{\rho}(\sigma),
\end{align}
and the multiplicity of the trivial irrep in the representation $\rho$ is
\begin{align}\label{multtriv}
	m_{\text{triv}}=  \frac{1}{|G|}\sum_{\sigma\in G} \chi(\sigma).
\end{align}

\subsection{Cosets, fast Fourier transform and hidden subgroups}
\label{sec:coset}

	If $G$ has a subgroup $H$ with $|H|$ elements, we can apply a group element $\sigma\in G$ to every element $\tau$ of $H$ to generate a left \textit{coset}
	\begin{align}
		\sigma H = \{\sigma \circ \tau | \tau \in H\}.
	\end{align}
	The group $G$ can be partitioned into $|G|/|H|$ distinct left cosets $\sigma_i H$, with $i=1,\dots |G|/|H|$,  where the $\sigma_i \in G$ are called coset \textit{representatives}.
	In other words, for every element $\sigma\in G$ there is a unique $i\in \{1,\dots |G|/|H|\}$ and a unique $\tau\in H$ such that $\sigma=\sigma_i\circ \tau$. The set of left cosets of $H$ in $G$ is denoted by $G/H$.
	Analogously, one can also partition $G$ into right cosets
	\begin{align}
	 H	\sigma = \{ \tau \circ \sigma  | \tau \in H\}.
	\end{align}

	We can use the decomposition of $G$ into left cosets to write the Fourier transform Eq.~\eqref{FT} as
	\begin{align}\label{lcosetdecomp}
		\hat{f}(\rho)&=	\sum_{\sigma\in G/H} \hat{\rho}(\sigma)\left(  \sum_{\tau\in H} f(\sigma\circ\tau) \hat{\rho}(\tau)\right),
	\end{align}
	where we write (slightly abusively) $\sigma\in G/H$ to signify that $\sigma$ runs over the set of representatives $\sigma_i$ of the cosets.
	The inner sum in Eq.~\eqref{lcosetdecomp} 
	is the Fourier transform over $H$ of a shifted version of $f$ at the representation $\rho\big|_H$ obtained by restricting $\rho$ to $H$. In this way, the Fourier transform can be iteratively evaluated on representations of smaller groups, generalizing the well-known Fast Fourier Transform algorithm for the efficient evaluation of the discrete Fourier transform \cite{clausen_fast_1989,diaconis_efficient_1990}.

In our upcoming investigations of the generalized Pauli principle (in Sec.~\ref{sec:Pauli}) and of completely destructive many-body interference (in Sec.~\ref{sec:suppr}), we will often encounter functions which are constant on each coset of a subgroup $H$ in $G$. In the field of quantum computing, such functions are said to \textit{hide the subgroup} $H$ \cite{kitaev_quantum_1995,jozsa_quantum_2001} (in that context, but not here, one additionally requires that the function take distinct values on the different cosets).
Given an arbitrary function $f$ on $G$, we obtain a function which hides a subgroup $H$ of $G$ by convoluting it with the normalized indicator function $I_H$, defined by
\begin{align}\label{indicator}
	I_H(\sigma)=\frac{1}{|H|} \quad \text{if } \sigma\in H \quad \text{and} \quad 
	I_H(\sigma)=\	0 \quad \text{otherwise.}
\end{align}
Indeed, for all $\sigma\in G$, we have
\begin{align}
	(f * I_H)(\sigma)&=\frac{1}{|H|}\sum_{\tau\in H} f(\sigma\circ\tau^{-1})\notag\\&=  \frac{1}{|H|}\sum_{\pi\in \sigma H} f(\pi),
\end{align}
such that right-convolution with $I_H$ amounts to averaging on the left cosets of $H$.
We therefore have the following equivalent conditions for $f$ to be constant over the left cosets of $H$:
\begin{align}\label{invf}
	(i)& \qquad \forall \sigma \in G,\ \forall \tau \in H, \quad   f(\sigma\circ\tau) = f(\sigma),\\
	(ii)& \qquad \forall \tau \in H, \qquad  f * \delta_\tau = f,\\
	(iii)& \qquad  f * I_H=f.\label{fIH=f}
\end{align}
Similar conditions apply for $f$ to be constant on the right cosets of $H$.


 Let us now turn to the  Fourier transform $\hat{I}_H(\lambda)$ of the indicator function of a subgroup $H$ at an irrep $\lambda$ of $G$.
 Given that $I_H*I_H=I_H$, we have, by virtue of \eqref{convolFT}, $[\hat{I}_H(\lambda)]^2=\hat{I}_H(\lambda)$, such that $\hat{I}_H(\lambda)$ is a projector. 
 Actually, writing out 
\begin{align}
	\hat{I}_H(\lambda)=\frac{1}{|H|}\sum_{\sigma\in H} \hat{\rho}^{(\lambda)}(\sigma),
\end{align} 
 we recognize the projector on the trivial components of the restriction $\rho^{(\lambda)}\big|_H$ of the $\lambda$ irrep  of $G$ to $H$ [see Eq.~\eqref{projtriv}]:
 \begin{align}
 \hat{I}_H(\lambda)=\hat{P}^{(\mathrm{triv})}(\rho^{(\lambda)}\big|_H ).
 \end{align}
 In particular, $\hat{I}_H(\lambda)=0$  if the trivial irrep does not appear in the decomposition of $\rho^{(\lambda)}\big|_H$, i.e., if the multiplicity [see Eq.~\eqref{multtriv}]
 \begin{align}
 	m_\mathrm{triv}=\frac{1}{|H|}\sum_{\sigma\in H} \chi^{(\lambda)}(\sigma)
 \end{align} 
 is equal to zero.
Following the terminology introduced in \cite{moore_explicit_2006}, we then say that $\lambda$ is a \textit{missing harmonic} of $H$.

From the above considerations, we obtain the following useful result: \textit{If a function $f$ hides a subgroup $H$ and $\lambda$ is a missing harmonic of $H$, then the Fourier component $\hat{f}(\lambda)$ vanishes.} Indeed if $f$ is constant on the left cosets of $H$, we have, by Eqs.~\eqref{fIH=f} and \eqref{convolFT},  $\hat{f}(\lambda) \hat{I}_H(\lambda)=\hat{f}(\lambda)$. Moreover, if  the trivial irrep does not appear in the decomposition of $\rho^{(\lambda)}\big|_H$, then $\hat{I}_H(\lambda)=0$, such that $\hat{f}(\lambda)=0$ follows. A similar argument applies if $f$ is constant on the right cosets of $H$. 
The invariance properties of a function can thus be used to prove that specific components are missing from its Fourier spectrum.
We will use this result in the formulation of a generalized Pauli principle in Sec.~\ref{sec:Pauli} and in our description of totally destructive interference in Sec.~\ref{sec:suppr-Pauli}.

\section{Many-particle transition amplitudes}
\label{sec:a}

\subsection{Transition amplitude function}



We now return to the problem of many-body interference in systems of identical particles, which we had introduced in Sec.~\ref{sec:MBItoFourier}. In Eq.~\eqref{firstdefa}, we had defined the \textit{transition amplitude function} $a$, which associates to each permutation $\sigma$ in the symmetric group $S_N$ the complex amplitude
\begin{align}\label{transamp}
 a(\sigma)=	\braket{\bm{o}| \hat{R}(\sigma)^\dagger U^{\otimes N}|\bm{i}}.
\end{align}
Recall that the input state $\ket{\bm{i}}$ and the output state $\ket{\bm{o}}$ are tensor products of $N$ single-particle basis states from an $M$-dimensional Hilbert space $\mathcal{H}$ [see Eq.~\eqref{basisvec}].  The unitary operator $\hat{R}(\sigma)$ reorders the $N$-fold tensor product according to the permutation $\sigma$  [see Eq.~\eqref{eq:reprSN}]. The $N$ particles are subjected to a non-interacting unitary evolution $\mathcal{U}=U^{\otimes N}$, with $U\in U(M)$ [see Eq.~\eqref{reprU}]. 

At this point, let us note that most of the results of this work \textit{do not} rely on the dynamics being non-interacting, i.e., on the above factorization of the $N$-particle evolution operator $\mathcal{U}$. Indeed, the only requirement to apply our formalism is the commutation of $\mathcal{U}$ with all permutations $\hat{R}(\sigma)$. In other words, the evolution must not distinguish between the particles, but it can have them interact. Keeping this in mind, we nevertheless proceed with a non-interacting evolution to better connect to the existing literature in the field of photonic interference.

The function $a$ and its Fourier transform  over the symmetric group will be the main objects of this work. While both contain the same information, the latter has the double advantage that the symmetric group acts upon it in a transparent way [as expressed by Eq.~\eqref{actSn}] and that it allows for an analysis in terms of the irreps of $S_N$.
To illustrate these points, we go back to our calculation at the end of Sec.~\ref{sec:MBItoFourier}. In Eq.~\eqref{transampcd}, we had considered the transition amplitude between two symmetrized states
\begin{align}\label{symstates}
	\hat{c}(R)\ket{\bm{i}}&=\sum_{\sigma\in S_N} c(\sigma) \hat{R}(\sigma)\ket{\bm{i}}\notag \\
	  \text{and} \qquad
	\hat{d}(R)\ket{\bm{o}}&=\sum_{\sigma\in S_N} d(\sigma) \hat{R}(\sigma)\ket{\bm{o}},
\end{align}
where the complex functions $c$ and $d$ over $S_N$ define the coefficients of the superpositions of permuted states (we disregard normalization for the time being).  
In the expression \eqref{convol+inner} of this transition amplitude, we had recognized a convolution product [Eq.~\eqref{convol}] and an inner product [Eq.~\eqref{scalarprod}]. With the notation introduced in Eqs.~\eqref{convol} and \eqref{scalarprod}, we can write
\begin{align}
	\braket{\bm{o}|\hat{d}(R)^\dagger U^{\otimes N} \hat{c}(R)|\bm{i}}=(d,c*a).
\end{align}
Using the the Parseval-Plancherel identity [Eq.~\eqref{PPid}] and the expression for the  Fourier transform of a convolution product[ Eq.~\eqref{convolFT}], we arrive at
\begin{align}\label{transitionarb}
	\braket{\bm{o}|\hat{d}(R)^\dagger U^{\otimes N} \hat{c}(R)|\bm{i}}= \sum_\lambda \frac{d_\lambda}{N!} \Tr\left[\hat{d}(\lambda)^\dagger \hat{c}(\lambda) \hat{a}(\lambda)   \right],
\end{align}
where the sum is over the irreps of the symmetric group $S_N$. We have thus obtained a decomposition of the transition amplitude into contributions associated with different types of particle-exchange symmetry, which we now turn to.



\subsection{Irreps of the symmetric group}
\label{sec:irreps}

We briefly describe the irreducible representations $\lambda$ of the symmetric group $S_N$, which appear in the sum in Eq.~\eqref{transitionarb}.
Again, we refer the reader to \cite{hamermesh_group_1989,chen,sengupta} for more details. Irreps of $S_N$ are indexed by integer partitions of $N$, that is by lists $\lambda=(\lambda_1,\lambda_2,\dots \lambda_r)$, where the integers $\lambda_1\geq \lambda_2 \dots \geq \lambda_r>0$ are such that $\sum_{k=1}^r\lambda_k=N$. These are conveniently represented by Young diagrams with $\lambda_i$ boxes in the $i$th row. For example, the Young diagrams associated with the irreps of $S_4$ are
\begin{equation}
	\ytableausetup{boxsize=0.75em}
	\begin{array}{ c c  c c c c }
	\lambda= &	(4)  & (3,1) &  (2,2) & (2,1,1) & (1,1,1,1)\\[2ex]
	 &\	\ydiagram{4}\ &\ \ydiagram{3,1}\ &\ \ydiagram{2,2}\ &\ \ydiagram{2,1,1}\ & \ \ydiagram{1,1,1,1}\
	\end{array}
\end{equation}
The dimension $d_\lambda$ of irrep $\lambda$ is given by the so-called hook-length formula \cite{chen}. It is equal to the number of standard Young tableaux of shape $\lambda$. These are obtained by filling the Young diagram with the integers $1, 2\dots N$ in such a way that the entries in each row and column (strictly) increase from left to right and from top to bottom. 
For the diagrams consisting of a single row or column, there is only one such filling, such that the corresponding irreps are one-dimensional: 
\begin{itemize}
	\item The \textit{trivial irrep}, associated with the partition $\lambda=(N)$, corresponding to the diagram with a single row, is such that for all $\sigma\in S_N$,   $\hat{\rho}^{(N)}(\sigma)=\chi^{(N)}(\sigma)=1$. 
	
	\item The \textit{sign irrep}, associated with the partition $\lambda=(1,\dots1)$, corresponding to the diagram with a single column, is such that  for all $\sigma\in S_N$,   $\hat{\rho}^{(1,\dots1)}(\sigma)=\chi^{(1,\dots1)}(\sigma)=\sign(\sigma)$.
\end{itemize}
Another prominent irrep is the $(N-1)$-dimensional \textit{standard irrep} (recall Tab.~\ref{tab:irrepsS3} for $S_3$), which is associated with the 
partition $\lambda=(N-1,1)$.

 Taking the tensor product of an irrep $\lambda$ with the sign irrep [i.e., multiplying the irrep matrices $\hat{\rho}^{(\lambda)}(\sigma)$ by $\sign(\sigma)$] yields its \textit{conjugate} irrep $\overline{\lambda}$, whose Young diagram is obtained from $\lambda$ by transposition. For example, the trivial and sign irreps are conjugate and the irrep $\lambda=(2, 1 \dots ,1)$ is conjugate to the standard irrep. In Young's orthogonal representation of $S_N$, the irrep matrices $\hat{\rho}^{(\lambda)}(\sigma)$ are real, orthogonal matrices \cite{kondor_group_2008}.
 The symmetric group's irreducible characters $\chi^{(\lambda)}(\sigma)$  [Eq.~\eqref{character}] are therefore real and obey $\chi^{(\lambda)}(\sigma^{-1})=\chi^{(\lambda)}(\sigma)$.


We are now ready to make our description of the symmetrization procedure outlined in Sec.~\ref{sec:MBItoFourier} more precise. 
Indeed, natural candidates for the symmetrization operators $\hat{c}(R)$ [Eq.~\eqref{eq:hatc}] are the projectors $\hat{P}^{(\lambda)}(R)$ on irreducible subrepresentations of type $\lambda$ of the representation $R$, as defined in Eq.~\eqref{isotyp}.
For the trivial irrep $\lambda=(N)$, we obtain the symmetrizer
\begin{align}\label{symmetrizer}
	\hat{P}^{(N)}(R)=\frac{1}{N!} \sum_{\sigma\in S_N} \hat{R}(\sigma),
\end{align}
which projects onto completely symmetric $N$-particle states, i.e., states of $N$ bosons.
For the sign irrep $\lambda=(1,\dots1)$, we get the antisymmetrizer
\begin{align}\label{asymmetrizer}
	\hat{P}^{(1,\dots1)}(R)=\frac{1}{N!} \sum_{\sigma\in S_N} \sign(\sigma) \hat{R}(\sigma),
\end{align}
which projects onto completely antisymmetric $N$-particle states, i.e., states of $N$ fermions.	
It is now natural to also consider many-body states which are obtained by applying the projectors $\hat{P}^{(\lambda)}(R)$ for irreps $\lambda$ beyond the trivial and sign irreps.
These have a particle-exchange symmetry which is neither totally symmetric nor totally antisymmetric \cite{messiah_symmetrization_1964,tichy_extending_2017}, so we will talk of a \emph{mixed symmetry}.
Note that while mixed symmetries generalize those of bosons and fermions, they are distinct from anyonic symmetries, which correspond to irreps of the braiding group \cite{kauffman_topology_2009}, rather than those of the symmetric group.  
Mixed-symmetry states are conceptually closer to the notion of paraparticles, particles whose states transform according to specific subsets of irreps $\lambda$ of $S_N$ \cite{hartle_quantum_1969,stolt_correspondence_1970}. 	
While they do not seem to be realized at an elementary level in nature,  
many-body states with a mixed symmetry can be created by suitably entangling distinguishable quantum systems.
For example, the states $\ket{\Psi_i^{(0)}}$ listed in Eq.~\eqref{3psymstates} can be viewed as entangled states of three qutrits with symmetry $\lambda=(2,1)$. Since this exchange symmetry is preserved under any evolution which does not distinguish between the subsystems, it provides an encoding of quantum information which is robust against global noise sources.
Moreover, as we will discuss in detail in Sec.~\ref{sec:J}, mixed symmetries are relevant for the description of systems of partially distinguishable bosons or fermions.

\subsection{Permanent, determinant and immanants}

\label{sec:permdetimm}

Let us now compute the transition amplitude Eq.~\eqref{transitionarb} in the case where the symmetrization operators $\hat{c}(R)$ and $\hat{d}(R)$ are projectors $\hat{P}^{(\lambda)}(R)$ and $\hat{P}^{(\mu)}(R)$ on irreps of type $\lambda$ and $\mu$, respectively.
Using Eq.~\eqref{hatPmu} in \eqref{transitionarb}, we immediately find
\begin{align}\label{immtransamp}
	\braket{\bm{o}| \hat{P}^{(\mu)}(R)^\dagger U^{\otimes N}  \hat{P}^{(\lambda)}(R)|\bm{i}}=\delta_{\lambda,\mu} 
	 \frac{d_\lambda}{N!} \Tr\left[ \hat{a}(\lambda)\right].
\end{align}
As expected, the transition amplitude must vanish if the input and output states do not share the same symmetry. If both states have symmetry $\lambda$, the transition amplitude is proportional to the trace of the Fourier transform $\hat{a}(\lambda)$ of the transition amplitude function \eqref{transamp} at irrep $\lambda$.
For a non-interacting evolution, this trace can be written as a matrix \emph{immanant}. The $\lambda$-immanant of an $N\times N$ matrix $\mathcal{M}$ is defined by \cite{littlewood_group_1934}
\begin{align}\label{eq:immanant}
	\text{imm}^\lambda(\mathcal{M})= \sum_{\sigma\in S_N} \chi^{(\lambda)}(\sigma)  \prod_{\alpha=1}^N \mathcal{M}_{\alpha,\sigma(\alpha)}.
\end{align}
Writing out $\hat{a}(\lambda)$ according to the definition \eqref{FT} of the Fourier transform, using the linearity of the trace and the definition \eqref{character} of the characters, we have
\begin{align}
	\Tr\left[ \hat{a}(\lambda)\right] &=\Tr\left[ \sum_{\sigma\in S_N} a(\sigma) \hat{\rho}^{(\lambda)}(\sigma)\right] \notag\\
	&=\sum_{\sigma\in S_N} a(\sigma) \chi^{(\lambda)}(\sigma).
\end{align}
Moreover, in the non-interacting case, the transition amplitudes \eqref{transamp} factorize as
\begin{align}
 a(\sigma)=\prod_{\alpha=1}^N \braket{o_{\sigma^{-1}(\alpha)} |U| i_\alpha }=\prod_{\alpha=1}^N \braket{o_\alpha |U| i_{\sigma(\alpha)}}.
\end{align}
We therefore have 
\begin{align}\label{imm}
	\Tr\left[ \hat{a}(\lambda)\right]  =\imm^{(\lambda)} \mathcal{A},
\end{align}
where the scattering matrix $\mathcal{A}$ has entries $\mathcal{A}_{\alpha,\beta}=\braket{o_\alpha|U|i_\beta}$, for  $\alpha,\beta=1,\dots N$.
For the trivial [$\lambda=(N)$] and sign [$\lambda=(1,\dots1)$] irreps, the immanant coincides with the permanent and determinant, respectively, and we immediately recover well-known results for bosonic  and fermionic transition amplitudes:
\begin{align}
	\braket{\bm{o}|\hat{P}^{(N)}(R)^\dagger  U^{\otimes N} \hat{P}^{(N)}(R)|\bm{i}}&=\frac{1}{N!} \hat{a}(N)\notag\\ &= \frac{1}{N!}\perm \mathcal{A},
\end{align}
\begin{align}	
	\braket{\bm{o}|\hat{P}^{(1,\dots1)}(R)^\dagger U^{\otimes N}\hat{P}^{(1,\dots1)}(R)|\bm{i}}&=\frac{1}{N!} \hat{a}(1,\dots1)\notag \\ &=\frac{1}{N!}\det \mathcal{A}.
\end{align}

While bosonic and fermionic states can be uniquely defined by specifying the number of particles occupying each mode, this is no longer the case for states with a mixed symmetry.
In particular, applying the projector $\hat{P}^{(\lambda)}(R)$ to a basis state $\ket{\bm{m}}$, to a permuted state  $\hat{R}(\sigma)\ket{\bm{m}}$ [see Eq.~\eqref{eq:reprSN}], or to a superposition $\hat{c}(R)\ket{\bm{m}}$ [see Eq.~\eqref{symstates}] may yield different results \cite{tichy_extending_2017}.
In particular, the  transition amplitude between the states
$\hat{P}^{(\lambda)}(R) \hat{c}(R) \ket{\bm{i}}$ and $\hat{P}^{(\lambda)}(R) \hat{d}(R) \ket{\bm{o}}$
 is given by a single term of the sum in Eq.~\eqref{transitionarb}:
 	\begin{align}\label{transarblambda}
 	&\braket{\bm{o}|\hat{d}(R)^\dagger \hat{P}^{(\lambda)}(R)^\dagger U^{\otimes N}\hat{P}^{(\lambda)}(R) \hat{c}(R) |\bm{i}}\notag\\ &\qquad \qquad=\frac{d_\lambda}{N!} \Tr\left[\hat{d}(\lambda)^\dagger\hat{c}(\lambda) \hat{a}(\lambda)\right],
 \end{align}
which is different from Eq.~\eqref{immtransamp} in general, and may not always be written as an immanant.



\subsection{Generalized Pauli principle}
\label{sec:Pauli}
Applying the symmetrizer $\hat{P}^{(N)}(R)$ [Eq.~\eqref{symmetrizer}] to an arbitrary basis state  $\ket{\bm{m}}$ always gives a non-zero result. In the case of the antisymmetrizer $\hat{P}^{(1,\dots1)}(R)$ [Eq.~\eqref{asymmetrizer}], on the other hand, it is well known that  Pauli's exclusion principle applies: $\hat{P}^{(1,\dots1)}(R) \ket{\bm{m}}$ vanishes as soon as a given mode appears more than once in $\ket{\bm{m}}$. For mixed symmetries,
the condition $\hat{P}^{(\lambda)}(R) \ket{\bm{m}}\neq 0$ enforces a \emph{generalized} Pauli principle \cite{tichy_extending_2017}:
\emph{No more than $\lambda_1$ particles can occupy the same mode, no more than $\lambda_1+\lambda_2$ particles can occupy the same pair of modes, and so on.}
In particular, for $\lambda=(N)$, there is no restriction on the mode occupations, while for $\lambda=(1,\dots1)$, no mode can be doubly occupied. 

The generalized Pauli principle is a consequence of Gamas' theorem \cite{gamas_conditions_1988,pate_immanants_1990,berget_short_2009}, which gives a necessary and sufficient condition for $\hat{P}^{(\lambda)}(R) \ket{\bm{m}}\neq 0$ \footnote{In full generality, Gamas' theorem states that for (not necessarily orthogonal) single-particle states $\ket{\varphi_1},\dots \ket{\varphi_N}$, 
	$\hat{P}^{(\lambda)}(R) \ket{\varphi_1,\dots \varphi_N}$ is non vanishing if and only if the $\ket{\varphi_i}$ can be partitioned into linearly independent sets whose sizes are given by the lengths of the columns of the Young diagram $\lambda$.}: It must be possible to write the mode labels $m_\alpha$ appearing in state $\ket{\bm{m}}$ in the boxes of the Young diagram $\lambda$ such that no label appears twice in the same column. This condition can be seen to coincide with our above statement of the generalized Pauli principle.
	 As an example, for $N=4$ and $\lambda=(2,1,1)$, $\hat{P}^{(\lambda)}(R) \ket{0,0,1,2}\neq 0$  since we can fill the corresponding Young diagram as follows
\begin{equation}
	\ytableausetup{boxsize=1em}\ytableaushort{00,1,2}
\end{equation}
but $\hat{P}^{(\lambda)}(R) \ket{0,0,1,1}=0$ since a total of four particles occupy modes $0$ and $1$, which is more than $\lambda_1+\lambda_2=3$, such that it is impossible to fill the first column of the Young diagram without repeating one of the mode labels.

Now that we know when it is satisfied, let us reformulate the condition 
$\hat{P}^{(\lambda)}(R) \ket{\bm{m}}\neq 0$ in the language of subgroup harmonics that we have introduced in Sec.~\ref{sec:coset}. 
Successively using the properties \eqref{propproj} and definition \eqref{isotyp} of the projectors $\hat{P}^{(\lambda)}(R)$, we can write 
\begin{align}
	\left\| \hat{P}^{(\lambda)}(R) \ket{\bm{m}}\right\| ^2&= \braket{\bm{m}| \hat{P}^{(\lambda)}(R) |\bm{m}}\notag\\
	&=  \frac{d_\lambda}{N!}\sum_{\sigma\in S_N} \chi^{(\lambda)}(\sigma) \braket{\bm{m}|\hat{R}(\sigma) |\bm{m}}.
\end{align}
The term $\braket{\bm{m}|\hat{R}(\sigma) |\bm{m}}$ vanishes unless $\sigma$ belongs to the \textit{stabilizer} of $\ket{\bm{m}}$:
\begin{align}\label{stab}
	\stab(\bm{m}) = \{  \sigma\in S_N : \hat{R}(\sigma)\ket{\bm{m}}=\ket{\bm{m}}  \}.
\end{align}
The stabilizer is a so-called Young subgroup of $S_N$, consisting of all permutations of particles which occupy the same mode. Its  cardinality is therefore 
\begin{align}
	|\stab(\bm{m})|= \prod_{m=0}^{M-1} n_m!,
\end{align}
 where $n_m$ is the number of times mode $m$ appears in $\bm{m}$. 
We can now rewrite the norm of $\hat{P}^{(\lambda)}(R) \ket{\bm{m}}$ as
\begin{align}
	\left\| \hat{P}^{(\lambda)}(R) \ket{\bm{m}}\right\| ^2= \frac{d_\lambda}{N!} \sum_{\sigma\in\stab(\bm{m})} \chi^{(\lambda)}(\sigma).
\end{align}
The above sum of characters gives the multiplicity \eqref{multtriv} of the trivial irrep  in the restriction $\rho^{(\lambda)}\big|_{\stab(\bm{m})}$ of $\lambda$ to $\stab(\bm{m})$.
We therefore have $\hat{P}^{(\lambda)}(R) \ket{\bm{m}}= 0$ if and only if $\lambda$ is a missing harmonic of $\stab(\bm{m})$.
This condition can also be expressed in terms of the Fourier transform of the indicator function of $\stab(\bm{m})$ (recall the end of Sec.~\ref{sec:coset}).
In the following, for the sake of brevity, we write $I_{\bm{m}}$ for $I_{\stab(\bm{m})}$ [see Eq.~\eqref{indicator}].
We then have $\hat{P}^{(\lambda)}(R) \ket{\bm{m}}= 0$ if and only if $\hat{I}_{\bm{m}}(\lambda)=0$.

The transition amplitude function $a$ between states $\ket{\bm{i}}$ and  $\ket{\bm{o}}$ [Eq.~\eqref{transamp}] satisfies
\begin{align}\label{Pauliinva}
	&\forall \tau\in \stab(\bm{o}), \ \forall \pi\in \stab(\bm{i}),\ \forall \sigma\in S_N,\notag \\ &\qquad\qquad  a(\tau\circ\sigma\circ\pi)=a(\sigma).
\end{align}
It follows that  
\begin{align}
	a &= I_{\bm{i}} * a * I_{\bm{o}}\\
	\intertext{and, by twofold application of \eqref{convolFT},}
	\hat{a}(\lambda) &= \hat{I}_{\bm{i}}(\lambda)  \hat{a}(\lambda)  \hat{I}_{\bm{o}}(\lambda),\label{inva}
\end{align}
for all irreps $\lambda$.
Therefore $\hat{a}(\lambda)$ vanishes if  $\hat{I}_{\bm{i}}(\lambda)=0$ or  $\hat{I}_{\bm{o}}(\lambda)=0$, i.e., if either $\ket{\bm{i}}$ or  $\ket{\bm{o}}$ does not comply with the generalized Pauli principle for irrep $\lambda$.
Such forbidden combinations of $\ket{\bm{i}}$, $\ket{\bm{o}}$ and $\lambda$ are marked in blue in Figs~\ref{FourierN4M4}-\ref{symsuppr}.

\section{Counting statistics in many-body interference}
\label{sec:countstat}

Rather than  transition amplitudes, many-body interference experiments typically measure counting statistics, i.e., the probability distribution over events where a given number of particles is found in each output mode, or marginals thereof.
Specifically, we want to evaluate the  probability $\mathrm{prob}(n_0,\dots n_{M-1})$ of measuring $n_0$ particles in output mode $0$, $n_1$ in mode $1$, and so on, up to $n_{M-1}$ in mode $M-1$. To do so, we define the following state
\begin{align}\label{orderedstate}
	\ket{\bm{o}}=\ket{\underbrace{0,\dots 0}_{n_0}, \underbrace{1, \dots 1}_{n_1} \dots ,\underbrace{M-1 \dots }_{n_{M-1}}}
\end{align}
with the desired occupations and form the projector on all of its distinct permutations (normalizing by the number of elements in $\stab(\bm{o})$ avoids double-counting):
\begin{align}\label{measurement}
	P_{\bm{o}}&=\frac{1}{|\stab(\bm{o})|} \sum_{\sigma\in S_N}   \hat{R}(\sigma) \ket{\bm{o}}\bra{\bm{o}} \hat{R}(\sigma)^\dagger, 
\end{align}
The expectation value of $P_{\bm{o}}$ in the output state yields the desired probability.
 In the following, we express such probabilities, for particles obeying different types of statistics, in terms of the Fourier transform of the transition amplitude function \eqref{transamp}. In particular, we consider the case of partially distinguishable bosons and fermions, and we will see that their states can also be submitted to a Fourier analysis.

 \subsection{Bosons, fermions, mixed symmetries and distinguishable particles}
 
 We start with an arbitrary (not necessarily normalized) initial state with input mode occupations specified by the mode list $\bm{i}$, which we again write as $\hat{c}(R)\ket{\bm{i}}$, for some function $c$ on $S_N$ [see Eq.~\eqref{symstates}].
 After a non-interacting evolution, the probability $\mathrm{prob}(n_0,\dots n_{M-1})$  is given by 
 \begin{align}\label{transprob}
 	\mathrm{prob}(n_0,\dots n_{M-1})=	\frac{\braket{\bm{i}|\hat{c}(R)^\dagger U^{\dagger \otimes N} P_{\bm{o}}  U^{\otimes N} \hat{c}(R) |\bm{i}}}{\braket{\bm{i}|\hat{c}(R)^\dagger \hat{c}(R)|\bm{i}}},
 \end{align}	
with $P_{\bm{o}}$ from Eq.~\eqref{measurement}. With the help of the Parseval-Plancherel relation \eqref{PPid}, we can express the numerator and denominator in terms of Fourier transforms [see Appendix.~\ref{app:cs}], yielding
\begin{widetext}
 	 \begin{align}\label{countstats}
 		 &\mathrm{prob}(n_0,\dots n_{M-1})=\frac{1}{|\stab(\bm{i})||\stab(\bm{o})|} \frac{\sum_\lambda d_\lambda\Tr\left[  \hat{a}(\lambda)^\dagger \hat{c}(\lambda)^\dagger \hat{c}(\lambda) \hat{a}(\lambda) \right]}
 		 	{ \sum_\lambda d_\lambda\Tr\left[  \hat{c}(\lambda)^\dagger \hat{c}(\lambda) \hat{I}_{\bm{i}}(\lambda) \right]}. 
 	\end{align}	
 We obtain the counting statistics of many-body states with symmetry $\lambda$  (including the bosonic and fermionic cases) by setting $\hat{c}(R)=\hat{P}^{(\lambda)}(R)\hat{d}(R)$, for an arbitrary symmetrization operator $\hat{d}(R)$. This has the effect of singling out one term in the sums over irreps, yielding 
\begin{align}\label{transproblambda}
 &\mathrm{prob}(n_0,\dots n_{M-1})=\frac{1}{|\stab(\bm{i})||\stab(\bm{o})|} \frac{\Tr\left[  \hat{a}(\lambda)^\dagger \hat{d}(\lambda)^\dagger \hat{d}(\lambda) \hat{a}(\lambda) \right]}
 { \Tr\left[  \hat{d}(\lambda)^\dagger \hat{d}(\lambda) \hat{I}_{\bm{i}}(\lambda) \right]}. 
\end{align}	
In particular, if $\hat{c}(R)=\hat{P}^{(\lambda)}(R)$, we find
\begin{align}
	 \mathrm{prob}(n_0,\dots n_{M-1})=\frac{1}{|\stab(\bm{i})||\stab(\bm{o})|} \frac{\Tr\left[  \hat{a}(\lambda)^\dagger  \hat{a}(\lambda) \right]}
	{ \Tr\left[ \hat{I}_{\bm{i}}(\lambda) \right]},
\end{align}	
which is essentially the spectral density \eqref{spectdens} of $a$. 
\end{widetext}

To obtain the counting statistics of distinguishable particles, we do not apply any symmetrization to the input state $\ket{\bm{i}}$ and thus take $c=\delta_{\id}$ in Eqs.~\eqref{transprob} and \eqref{countstats}, yielding
\begin{align}\label{CSdist}
 \mathrm{prob}(n_0,\dots n_{M-1})&=	\braket{\bm{i}|U^{\dagger \otimes N} P_{\bm{o}}  U^{\otimes N} |\bm{i}}\\
	&=\frac{1}{|\stab(\bm{o})|}\sum_\lambda \frac{d_\lambda}{N!}\Tr\left[  \hat{a}(\lambda)^\dagger \hat{a}(\lambda) \right]. \notag
\end{align}	
Using the Parseval-Plancherel identity \eqref{PPid}, we can rewrite the above transition probability as 
\begin{align}
 \mathrm{prob}(n_0,\dots n_{M-1})&=\frac{1}{|\stab(\bm{o})|} (a,a)\notag \\
 &=\frac{1}{|\stab(\bm{o})|} \sum_{\sigma\in S_N} |a(\sigma)|^2\notag\\
 &=\frac{1}{|\stab(\bm{o})|}\perm|\mathcal{A}|^2,
\end{align} 
where $|\mathcal{A}|^2$  is the matrix with elements $|\braket{o_\alpha|U|i_\beta}|^2$, $\alpha,\beta=1,\dots N$. We thus recover the fact that the transition probabilities of distinguishable particles can be expressed as permanents of matrices with positive entries \cite{tichy_sampling_2015}.

\subsection{Partially distinguishable bosons and fermions}
\label{sec:J}

While particles obeying a mixed exchange symmetry remain rather hypothetical objects, mixed symmetries are central to a topic of great practical relevance: the impact of partial distinguishability on the counting statistics of bosons or fermions.
Partially distinguishable particles posses internal states which act as (possibly ambiguous) labels: 
if they are distinct enough, they allow  to identify the particles, many-body paths become distinguishable from one another and many-body interference is suppressed; if they are similar, they do not unequivocally individuate the particles and interference is restored \cite{dittel_waveparticle_2021}.
To describe partially distinguishable particles, we extend the single-particle Hilbert space $\mathcal{H}$ (hereafter, \textit{external} Hilbert space) by taking its tensor product with an \emph{internal} state space $\mathcal{I}$.
 The $N$-particle Hilbert space is thus $\left( \mathcal{H} \otimes \mathcal{I}\right)^{\otimes N}$, but it is convenient to group the external and internal tensor factors as $\mathcal{H}^{\otimes N} \otimes \mathcal{I}^{\otimes N}$. 
 We then consider bosonic or fermionic states on this $N$-particle Hilbert space, but assume that only 
 the particles' external degrees of freedom evolve and are ultimately measured, such that we can  trace out the internal Hilbert spaces $\mathcal{I}^{\otimes N}$ and work with the reduced states on $\mathcal{H}^{\otimes N}$.
  
  In the full Hilbert space $\mathcal{H}^{\otimes N} \otimes \mathcal{I}^{\otimes N}$, permutations of the particles correspond to joint permutations of the $N$-particle external and internal states. 
  These are performed by the representation $R\otimes R_\mathrm{int}$, where $R$ permutes the factors of  $\mathcal{H}^{\otimes N}$ [recall Eq.~\eqref{eq:reprSN}] and $R_\mathrm{int}$ acts analogously on $\mathcal{I}^{\otimes N}$.
In particular, the symmetrizer $\hat{P}^{(N)}(R\otimes R_\mathrm{int})$ and the antisymmetrizer $\hat{P}^{(1,\dots1)}(R\otimes R_\mathrm{int})$ are given by (we introduce a unified notation for convenience)
 \begin{align}\label{symasym}
 	\hat{S}_\epsilon= \frac{1}{N!}\sum_{\sigma\in S_N} \epsilon(\sigma) \hat{R}(\sigma) \otimes \hat{R}_\mathrm{int}(\sigma),
 \end{align}
 with either $\epsilon(\sigma)=1$ for all $\sigma\in S_N$, in which case $\hat{S}_\epsilon=\hat{P}^{(N)}(R\otimes R_\mathrm{int})$ is the symmetrizer [compare Eq.~\eqref{symmetrizer}], or $\epsilon(\sigma)=\sign(\sigma)$ for all $\sigma\in S_N$, in which case $\hat{S}_\epsilon=\hat{P}^{(1,\dots1)}(R\otimes R_\mathrm{int})$ is the antisymmetrizer [compare Eq.~\eqref{asymmetrizer}].
 General (possibly mixed) bosonic and fermionic $N$-particle states $\varrho_\mathrm{tot}$ on  $\mathcal{H}^{\otimes N} \otimes \mathcal{I}^{\otimes N}$ are defined by the condition
 	\begin{align}
	 \varrho_\mathrm{tot}&= \hat{S}_\epsilon\varrho_\mathrm{tot}\hat{S}_\epsilon.
\end{align}	
   In particular, for both bosonic and fermionic states, we have, $\forall \sigma\in S_N$,
 	\begin{align}\label{invariance}
 	 \hat{R}(\sigma) \otimes \hat{R}_\mathrm{int}(\sigma) \varrho_\mathrm{tot} \hat{R}(\sigma)^\dagger \otimes \hat{R}_\mathrm{int}(\sigma)^\dagger =\varrho_\mathrm{tot}.
 	\end{align}
 	Taking the trace over internal degrees of freedom on both sides of Eq.~\eqref{invariance}, we find that the external state
 	\begin{align}
 		\varrho =  \Tr_{\mathcal{I}^{\otimes N}}(\varrho_\mathrm{tot})
 	\end{align}
 	 satisfies
 	\begin{align}\label{statecommute}
 		\forall \sigma\in S_N, \qquad\hat{R}(\sigma) \varrho \hat{R}(\sigma)^\dagger = \varrho,
 	\end{align}
 	i.e., $\varrho$ commutes with all unitary permutation operators $\hat{R}(\sigma)$.

	We now assume that the external mode occupations are fixed, such that $\varrho$ is supported on 	the subspace of $\mathcal{H}^{\otimes N}$ spanned by states of the form $\hat{R}(\sigma)\ket{\bm{i}}$ for a given mode list $\bm{i}$ and $\sigma$ running over $S_N$.
	All non-vanishing matrix elements of $\varrho$ are therefore of the form
	\begin{align}
		\braket{\bm{i}|\hat{R}(\sigma)^\dagger\varrho\hat{R}(\tau)|\bm{i}}=\braket{\bm{i}|\hat{R}(\tau^{-1}\circ\sigma)^\dagger\varrho|\bm{i}},
	\end{align} 
	for permutations $\sigma$ and $\tau$ in $S_N$. To derive the above equality, we have used  Eq.~\eqref{statecommute}, together with the fact that $R$ is a unitary representation.
	Complete information about the state $\varrho$ can thus be encoded in a \emph{partial distinguishability function} $j:S_N\to \mathbb{C}$, defined in analogy with the transition amplitude function of Eq.~\eqref{transamp} by
	\begin{align}\label{eq:j}
		j:\sigma\mapsto \frac{N!}{|\stab(\bm{i})|}\braket{\bm{i}|\hat{R}(\sigma)^\dagger \varrho  |\bm{i}}.
	\end{align}
 Given $j$, one can reconstruct $\varrho$ as
	\begin{align}\label{eq:rhoj}
		\varrho= \frac{1}{N!|\stab(\bm{i})|} \sum_{\sigma,\tau\in S_N} j(\tau^{-1}\circ\sigma) \hat{R}(\sigma)\ket{\bm{i}}\bra{\bm{i}}\hat{R}^\dagger(\tau).
	\end{align}

We are now ready to calculate transition probabilities for partially distinguishable particles whose reduced external state is initially $\varrho$ and which are submitted to a non-interacting evolution $U^{\otimes N}$ of the external degrees of freedom. 
The probability $\mathrm{prob}(n_0,\dots n_{M-1})$ of measuring final occupations $n_0, n_1, \dots n_{M-1}$ associated with a  mode list $\bm{o}$ [see Eq.~\eqref{orderedstate}] is again given by the expectation value of the projector $P_{\bm{o}}$:
\begin{align}
	\mathrm{prob}(n_0,\dots n_{M-1})=	\Tr(U^{\otimes N} \varrho U^{\dagger\otimes N} P_{\bm{o}}).
\end{align}
In Appendix.~\ref{app:cs}, we show that this probability can be expressed in terms of the Fourier transforms $\hat{j}(\lambda)$ of the partial distinguishability function and $\hat{a}(\lambda)$ of the transition amplitude function (see also Eq.~(3.4.126) of \cite{geller_characterization_2024}):
\begin{align}\label{countstat-partdist}
	&\mathrm{prob}(n_0,\dots n_{M-1})\\
	&\qquad =\frac{1}{|\stab(\bm{o})| |\stab(\bm{i})|} \sum_\lambda  \frac{d_\lambda}{N!} \Tr\left[\hat{a}(\lambda)^\dagger\hat{j}(\lambda)\hat{a}(\lambda)\right].\notag
\end{align}
The transition probabilities of partially distinguishable bosons or fermions therefore receive contributions from the various symmetry sectors $\lambda$, and not only the bosonic and fermionic ones.
Indeed, the overall bosonic or fermionic symmetry of the many-body state $\varrho_\mathrm{tot}$ is broken upon tracing out the internal degrees of freedom. 
The Fourier transform $\hat{j}(\lambda)$ of the partial distinguishability function encodes the transformation properties of the resulting reduced state $\varrho$ under particle exchange.
 As we show in Appendix~\ref{app:positivity}, the matrices $\hat{j}(\lambda)$ inherit the Hermiticity and positivity properties of the state $\varrho$, and can therefore also be viewed, up to normalization, as density matrices.
 
Quantities essential to the characterization of  the particles' partial distinguishability are conveniently expressed in terms of the  $\hat{j}(\lambda)$.
As shown in Appendix \ref{app:weights}, one can for example compute the purity of the state $\varrho$ as
\begin{align}\label{eq:purity}
	\Tr\left[  \varrho^2 \right]=\frac{1}{N!}\sum_\lambda \frac{d_\lambda}{N!} \Tr\left[ \hat{j}(\lambda)^2\right],
\end{align}
and its weight on symmetry sector $\lambda$ as
	\begin{align}\label{eq:weights}
		p_\lambda&=\Tr\left[  \hat{P}^{(\lambda)}(R)\varrho \right]=\frac{d_\lambda}{N!} \Tr[\hat{j}(\lambda)].
	\end{align}
The purity of the reduced external state $\varrho$ quantifies the remaining many-body coherence upon discarding the internal states and determines the magnitude of many-body interference contributions to the dynamics of the external degrees of freedom \cite{shchesnovich_partial_2015,dittel_waveparticle_2021}. 
In a number of works \cite{tichy_sampling_2015,shchesnovich_tight_2015,englbrecht_indistinguishability_2024}, 
the weight $p_{(N)}=\hat{j}[\lambda=(N)]/N!$ of the bosonic sector  is proposed as a measure of bosonic indistinguihsability. 
Arguably, the entire collection of weights \eqref{eq:weights}, for all irreps $\lambda$ of $S_N$, provides a more complete picture, and allows for a finer understanding of the impact of partial distinguishability on observable quantities \cite{brunner_manybody_2023}.
However, the matrices $\hat{j}(\lambda)$ do not only enter in Eq.~\eqref{countstat-partdist} through their trace \eqref{eq:weights}, and their individual matrix elements may also be relevant.
While the full Fourier spectrum $\hat{j}(\lambda)$, for all irreps $\lambda$ of $S_N$, contains the same  information as the collection of  $N!$ matrix elements $j(\sigma)$ of $\varrho$ [recall Eqs.~\eqref{eq:j} and \eqref{eq:rhoj}], one can, depending on the physical situation, achieve a more economical characterization of partial distinguishability by focusing only on a few relevant sectors $\lambda$.
For example, for bosons that are close to being indistinguishable, the single quantifier $\hat{j}[\lambda=(N)]$ can be complemented by the $(N-1)\times (N-1)$ matrix $\hat{j}[\lambda=(N-1,1)]$, which we expect to carry the most relevant information on deviations from perfect indistinguishability, given that $(N-1,1)$ is the partition  which is closest to $(N)$ (in the sense that the corresponding Young diagrams are related by displacing a single box).

We now examine how the properties of the reduced external state $\varrho$---as encoded in the partial distinguishability function $j$---relate to the internal state of the particles. To this end, we consider an arbitrary $N$-particle internal state described by a density operator $\varrho_\mathrm{int}$ on $\mathcal{I}^{\otimes N}$ and form the following bosonic or fermionic state whose external mode occupations are given by the mode list $\bm{i}$:
\begin{align}\label{eq:rhotot}
	\varrho_{\text{tot}}= \mathcal{N} \hat{S}_\epsilon \left(   \ket{\bm{i}}\bra{\bm{i}} \otimes \varrho_\mathrm{int} \right) \hat{S}_\epsilon.
\end{align}
Here, $\mathcal{N}$ is a normalization constant ensuring that $\Tr(\varrho_\text{tot})=1$, and $\hat{S}_\epsilon$ is the (anti)symmetrizer [Eq.~\eqref{symasym}]. 
Taking the partial trace  of $\varrho_{\text{tot}}$ over $\mathcal{I}^{\otimes N}$ yields the reduced external state $\varrho$. 
Actually, as we demonstrate in Appendix \ref{app:jJrhoint},  any reduced external state $\varrho$ with external mode occupations given by the mode list $\bm{i}$ and 
satisfying Eq.~\eqref{statecommute} can be obtained in this manner.

As also shown in Appendix \ref{app:jJrhoint}, the partial distinguishability function \eqref{eq:j} associated with the  reduced external state $\varrho$ can be written as
\begin{align}\label{eq:jJ}
	j=   \frac{I_{\bm{i}}* J * I_{\bm{i}}}{(I_{\bm{i}}, J)},
\end{align}
where $I_{\bm{i}}$ is the indicator function  of the stabilizer of $\ket{\bm{i}}$ [recall Eq.~\eqref{indicator}] and the function $J$ is defined by
\begin{align}\label{eq:J}
\forall \sigma\in S_N,\qquad	J(\sigma)= \epsilon(\sigma) \Tr\left[\varrho_\text{int}\hat{R}_\mathrm{int}(\sigma)  \right].
\end{align}
This function was already introduced in \cite{shchesnovich_sufficient_2014,shchesnovich_partial_2015,shchesnovich_tight_2015,shchesnovich_universality_2016}. It depends only on the internal state $\varrho_\mathrm{int}$ and --- through $\epsilon$--- on the quantum statistics of the particles.
In Eq.~\eqref{eq:jJ}, the dependence of $j$ on the external states of the particles enters through the indicator function $I_{\bm{i}}$. This dependence only comes about because (anti)symmetrizing the overall state $\varrho_{\mathrm{tot}}$ [Eq.~\eqref{eq:rhotot}] always (anti)symmetrizes the internal states of particles which occupy the same external mode.
Without loss of generality, we can thus take $\varrho_{\mathrm{int}}$ to already be (anti)symmetric with respect to the exchange of particles occupying the same mode, i.e., $\forall\tau\in \stab(\bm{i}),$
\begin{align}
	 \hat{R}_{\mathrm{int}}(\tau)\varrho_{\mathrm{int}}=\varrho_{\mathrm{int}}\hat{R}_{\mathrm{int}}(\tau)=\epsilon(\tau) \varrho_{\mathrm{int}}.
\end{align}
If this is the case, we have $I_{\bm{i}}*J=J*I_{\bm{i}}=J$ and $(I_{\bm{i}},J)=1$, such that Eq.~\eqref{eq:jJ} reduces to $j=J$. 
Moreover, for states with at most one particle per external mode, the stabilizer $\stab(\bm{i})=\{\id\}$ is trivial and we again have $j=J$.

With Eq.~\eqref{eq:jJ}, the transition probability \eqref{countstat-partdist} can be expressed directly in terms of the Fourier transform of $J$. Making use of Eqs.~\eqref{convolFT}, \eqref{PPid} and \eqref{inva}, we obtain
	\begin{align}\label{PJ}
		&\mathrm{prob}(n_0,\dots n_{M-1})\\
		&\qquad =\frac{1}{|\stab(\bm{o})| |\stab(\bm{i})|}   \frac{\sum_\lambda d_\lambda\Tr\left[\hat{a}(\lambda)^\dagger\hat{J}(\lambda)\hat{a}(\lambda)\right]}{\sum_\lambda d_\lambda\Tr\left[\hat{I}_{\bm{i}}(\lambda)\hat{J}(\lambda)\right]}.\notag
\end{align}
In the following, we examine the properties of the function $J$ and of its Fourier transform $\hat{J}(\lambda)$, and their dependence on the internal state $\varrho_{\mathrm{int}}$. 

Since
$\sign(\sigma)\hat{\rho}^{(\lambda)}(\sigma)=\hat{\rho}^{(\overline{\lambda})}(\sigma)$, with $\overline{\lambda}$ the partition conjugate to $\lambda$, the Fourier transforms of $J$ in the bosonic and fermionic cases are related by
\begin{align}\label{JbosJfer}
	\hat{J}_\text{bosons}(\lambda)=\hat{J}_\text{fermions}(\overline{\lambda}).
\end{align} 
This reflects the fact, known as unitary-unitary duality \cite{rowe_dual_2012}, that bosonic states are formed by combining internal and external states which share the same symmetry $\lambda$, whereas fermionic states are obtained by pairing internal and external states with conjugate symmetries $\lambda$ and $\overline{\lambda}$.

In the case where each particle $\alpha=1,\dots N$ is in a pure internal state $\ket{\varphi_\alpha}\in \mathcal{I}$, such that
 $\varrho_\text{int}=\ket{\psi_\text{int}}\bra{\psi_\text{int}}$, with $\ket{\psi_\text{int}}=\ket{\varphi_1}\otimes\ket{\varphi_2} \otimes\dots \ket{\varphi_N}$, we have 
\begin{align}
	J(\sigma) &=\epsilon(\sigma)
	\braket{\psi_\text{int}|  \hat{R}_\mathrm{int}(\sigma)  |\psi_\text{int}}\notag \\
	&= \epsilon(\sigma)
	\prod_{\alpha=1}^N \braket{\varphi_{\sigma(\alpha)}|\varphi_{\alpha}}\notag \\
	&=\epsilon(\sigma)
	\prod_{\alpha=1}^N S_{\sigma(\alpha),\alpha},
\end{align}
where the Gram matrix $S=( \braket{\varphi_\alpha|\varphi_\beta} )_{\alpha,\beta=1}^N$ is known as the distinguishability matrix \cite{tichy_sampling_2015}. Provided $J=j$ (e.g., for states with at most one particle per mode), the components $\hat{J}[\lambda=(N)]$ and $\hat{J}[\lambda=(1,\dots1)]$ determine the weights associated with the trivial and sign irreps [recall Eq.~\eqref{eq:weights}].
They are respectively given by the permanent and determinant of $S$ for bosons, and vice-versa for fermions [recall Eq.~\eqref{JbosJfer}]. 
 This supports the interpretation of $\perm S$ as a simple measure of indistinguishability \cite{tichy_sampling_2015,shchesnovich_partial_2015} in both the bosonic and fermionic cases. However, as discussed above, a finer characterization of indistinguishability should also take other components $\hat{J}(\lambda)$ into account.

We now further specialize to the situation  where the internal states $\ket{\varphi_\alpha}$ are either equal or orthogonal, such that any two particles are either perfectly indistinguishable or  perfectly distinguishable from one another. Let us thus take $\ket{\psi_\text{int}}=\ket{\bm{s}}=\ket{s_1}\otimes\ket{s_2} \otimes\dots \ket{s_N}$ where the $\ket{s_\alpha}, \, \alpha=1,\dots N,$ are picked from an orthonormal basis of the internal Hilbert space $\mathcal{I}$. We then have 
\begin{align}
J(\sigma) =\epsilon(\sigma)
	\braket{\bm{s}|  \hat{R}_\mathrm{int}(\sigma)  |\bm{s}}= |\stab(\bm{s})|\epsilon(\sigma)
 I_{\bm{s}}(\sigma),
\end{align}
with $I_{\bm{s}}$ the indicator function of the stabilizer of $\ket{\bm{s}}$ for the action $R_\text{int}$ of $S_N$. It follows that
\begin{align}
	\hat{J}(\lambda)=\begin{cases}
		|\stab(\bm{s})|\hat{I}_{\bm{s}}(\lambda) \quad\text{for bosons,}\\
	|\stab(\bm{s})|	\hat{I}_{\bm{s}}(\overline{\lambda}) \quad\text{for fermions.}
	\end{cases}  
\end{align}
Assuming $j=J$, the sector weights \eqref{eq:weights} are then given by
	\begin{align}
		p_\lambda
		&= \frac{d_\lambda}{N!} \sum_{\sigma \in \stab{\bm{s}}} \epsilon(\sigma) \chi^{(\lambda)}(\sigma).
	\end{align}
In this expression, we recognize the multiplicity \eqref{multtriv} of the trivial irrep in the restriction of the $\lambda$ irrep of $S_N$ (in the case of bosons) or of the $\overline{\lambda}$ irrep of $S_N$ (in the case of fermions) to $|\stab(\bm{s})|$.
These can be expressed in terms of Kostka numbers, which, for two partitions $\lambda$ and $\mu$, give the multiplicity of the trivial irrep in the restriction of the $\lambda$ irrep of $S_N$ to the Young subgroup  $S_\mu=S_{\mu_1}\times S_{\mu_2}\times\dots$ with $|S_\mu|=\mu_1!\mu_2!\dots$ elements \cite{fulton_representation_2004}:
\begin{align}
	K(\lambda, \mu) = \frac{1}{|S_\mu|}\sum_{\sigma \in S_\mu} \chi^{(\lambda)}(\sigma).
\end{align}
If $\mu(\bm{s})$ is the partition obtained by recording the number of times each internal basis state occurs in $\ket{\bm{s}}$, we have
	\begin{align}
		p_\lambda =  \begin{cases}
			d_\lambda\,|\stab(\bm{s})| K(\lambda, \mu(\bm{s}))/N!  \quad \text{for bosons,}\\
			d_\lambda\,|\stab(\bm{s})| K(\overline{\lambda},\mu(\bm{s}))/N!\quad \text{for fermions.}
		\end{cases}  
	\end{align}
The generalized Pauli principle described in Sec.~\ref{sec:Pauli} therefore also applies to internal states $\ket{\bm{s}}$: if the entries of $\bm{s}$ cannot be written in the Young diagram $\lambda$ without an internal state appearing twice in the same column, then $\hat{J}_\text{bosons}(\lambda)=\hat{J}_\text{fermions}(\overline{\lambda})=0$. 	
For a given internal state $\ket{\bm{s}}$, the generalized Pauli principle thus selects those sectors $\lambda$ which contribute to the transition probability \eqref{PJ}. 

\subsection{Pure state representation for partially distinguishable particles}

We conclude our discussion of partially distinguishable particles by noting an interesting consequence of the positivity of the matrices $\hat{J}(\lambda)$ (which follows from the positivity of $\hat{j}(\lambda)$, see Appendix \ref{app:positivity}, since the two are equal for states with at most one particle per mode):
For each irrep $\lambda$, we can find a matrix $\hat{c}(\lambda)$ such that $\hat{J}(\lambda)=\hat{c}(\lambda)^\dagger\hat{c}(\lambda)$ (see also \cite{shchesnovich_universality_2016} for the corresponding factorization of the function $J$). 
Applying the inverse Fourier transform \eqref{IFT} to the $\hat{c}(\lambda)$, we obtain a function $c$, which we use to define a symmetrization operator $\hat{c}(R)$ [recall Eq.~\eqref{eq:hatc}].
The expression \eqref{PJ} of the transition probability for partially distinguishable particles then coincides with the expression \eqref{countstats} of the transition probability for a pure initial state $\hat{c}(R)\ket{\bm{i}}$.
This is a rather remarkable result: The (pure) superposition state $\hat{c}(R)\ket{\bm{i}}$ generates the same counting statistics as the (in general mixed) reduced external state $\varrho$.

A relatively trivial illustration of this is the case of distinguishable particles.
If $\dim(\mathcal{I})\geq N$, we can assign a distinct internal state to each particle, which produces the reduced external state 
\begin{align}
   \varrho=\frac{1}{N!}\sum_{\sigma\in S_N} \hat{R}(\sigma)\ket{\bm{i}}\bra{\bm{i}}\hat{R}(\sigma)^\dagger,
\end{align}
a statistical mixture of all permutations of $\ket{\bm{i}}$.
In this case, we have $J=c=\delta_{\id}$ for both bosons and fermions, and the transition probability \eqref{PJ} is indeed found to coincide with Eq.~\eqref{CSdist}, which we had calculated for the pure initial state $\hat{c}(R)\ket{\bm{i}}=\ket{\bm{i}}$.
For a less obvious example, consider three bosons with non-orthogonal internal states $\ket{0}$ and $\frac{1}{2}\ket{0}\pm \frac{\sqrt{3}}{2}\ket{1}$, where $\ket{0}$ and $\ket{1}$ form a basis of a two-dimensional internal Hilbert space $\mathcal{I}$. In this case, we obtain
\begin{align}
	\hat{J}[\lambda=(3)]&= \frac{3}{2},\notag \\
	\hat{J}[\lambda=(2,1)]&=\frac{9}{8}\mathbb{I}_2, \\
	\hat{J}[\lambda=(1,1,1)]&=0.\notag 
\end{align}
Taking the square root and applying the inverse Fourier transform to obtain $c$, we find (choosing $\ket{\bm{i}}=\ket{0,1,2}$) that the pure state
\begin{align}
	\hat{c}(R)\ket{0,1,2}&=\frac{1}{2 \sqrt{6}}\Big[  (1+2\sqrt{3})\ket{0,1,2}+\ket{0,2,1}\notag\\
	&\qquad +\ket{1,0,2}+(1-\sqrt{3})\ket{2,0,1}\\
	&\qquad +(1-\sqrt{3})\ket{1,2,0}+\ket{2,1,0}\Big]\notag
\end{align}
reproduces the counting statistics of the three partially distinguishable bosons. Note that this state is conveniently expressed in the basis \eqref{3psymstates} as
\begin{align}
	\hat{c}(R)\ket{0,1,2}=\frac{1}{2}\ket{\psi^{(+)}}+\frac{\sqrt{3}}{2}\ket{\psi^{(0)}_1}.
\end{align}

\section{Completely destructive interference}
\label{sec:suppr}

As an exemplary application of the formalism developed in the previous sections, we study the mechanisms behind completely destructive many-body interference, i.e., exact cancellations of transition amplitudes between input and output states with a given symmetry. 
The suppression of coincidence events for bosons in the Hong-Ou-Mandel experiment \cite{hong_measurement_1987,bouchard_twophoton_2021}, which we touched upon in Sec.~\ref{sec:MBItoFourier}, is the most famous and simple example of such completely destructive many-body interference. 
Given the many fundamental and practical implications of the Hong-Ou-Mandel effect \cite{bouchard_twophoton_2021}, there have been efforts to generalize it to more particles in larger interferometers \cite{lim_generalized_2005,tichy_zerotransmission_2010,tichy_manyparticle_2012,tichy_stringent_2014,crespi_suppression_2015,crespi_suppression_2016,dittel_manybody_2017,dittel_totally_2018,dittel_totally_2018a}. Fermions have also been shown to display destructive many-body interference beyond the Pauli principle \cite{tichy_manyparticle_2012,dittel_totally_2018,dittel_totally_2018a}. Suppressions have been shown to stem from the invariance (up to a phase)  of the input state under a permutation of the interferometer modes, allowing for further generalizations to states 
 featuring partial distinguishability and/or entanglement \cite{dittel_totally_2018a,dittel_interference_2019}.
 Here we consider for the first time instances of totally destructive many-body interference for states which transform according to higher-dimensional irreps of the symmetric group.

Given an input state $\ket{\bm{i}}$, an output state $\ket{\bm{o}}$, and a unitary transformation $U$---defining a transition amplitude function $a$ [Eq.~\eqref{transamp}]---we say that the transition is suppressed in symmetry sector $\lambda$ if the Fourier transform of the transition amplitude function vanishes at irrep $\lambda$:
\begin{gather}
	\hat{a}(\lambda)=0,\\ \intertext{or, equivalently, in terms of the spectral density \eqref{spectdens},} \Tr[\hat{a}(\lambda)^\dagger \hat{a}(\lambda)]=0.
\end{gather}
In particular, this means that the transition probability \eqref{transproblambda} between states obtained by symmetrizing $\ket{\bm{i}}$ and $\ket{\bm{o}}$ such that they have exchange symmetry  $\lambda$ vanishes. Moreover, the transition probability \eqref{transprob} between states of the form \eqref{symstates}, obtained from $\ket{\bm{i}}$ and $\ket{\bm{o}}$ by applying arbitrary symmetrization operators, receives no contribution from the irrep $\lambda$. The same is true for the transition probability \eqref{countstat-partdist} of partially distinguishable bosons or fermions.
Note, however, that there may be situations where the contribution of the symmetry sector $\lambda$ to a transition probability vanishes although $\hat{a}(\lambda)\neq 0$. For example, for partially distinguishable particles, it can happen that $\Tr\left[\hat{a}(\lambda)^\dagger \hat{j}(\lambda) \hat{a}(\lambda) \right]=0$ although $\hat{a}(\lambda)\neq 0$.
In the following, we describe two mechanisms which lead to suppressed transitions in a given symmetry sector.

\subsection{Symmetry-induced suppressions}
\label{sec:suppr-sym}

The first type of suppression occurs when there exists a permutation $\tau\in S_N$ 
such that any two transition amplitudes related by a permutation $\hat{R}(\tau)$ of the input state $\ket{\bm{i}}$ differ by  a constant factor $\Lambda\in \mathbb{C}$: $\forall \sigma\in S_N,$
\begin{align}
	  \braket{ \bm{o} | \hat{R}(\sigma)^\dagger   U^{\otimes N}\hat{R}(\tau)|\bm{i}  }=\Lambda \braket{ \bm{o} | \hat{R}(\sigma)^\dagger   U^{\otimes N}|\bm{i}  }.
\end{align}
In terms of the transition amplitude function \eqref{transamp} and the shift operation \eqref{shift}, this means that 
\begin{align}\label{eq:supprsym}
	\delta_\tau * a= \Lambda a.
\end{align}
By Eq.~\eqref{actSn},  it follows that for all irreps $\lambda$,
\begin{align}
	\hat{\rho}^{(\lambda)}(\tau)\hat{a}(\lambda)=\Lambda\hat{a}(\lambda).
\end{align}
Therefore, the image of $\hat{a}(\lambda)$ lies in the eigenspace of $\hat{\rho}^{(\lambda)}(\tau)$  with eigenvalue $\Lambda$. We thus have the following suppression criterion: if $\Lambda$ is absent from the spectrum of $\hat{\rho}^{(\lambda)}(\tau)$, then $\hat{a}(\lambda)=0$.
For the one-dimensional bosonic and fermionic irreps, this condition immediately translates to:
\begin{itemize}
	\item if $\Lambda\neq 1$, the bosonic transition amplitude $\hat{a}[\lambda=(N)]$ must vanish, since $\hat{\rho}^{(N)}(\tau)=1$,

	\item if $\Lambda\neq \text{sign}(\tau)$, the fermionic transition amplitude $\hat{a}[\lambda=(1,\dots1)]$ must vanish, since $\hat{\rho}^{(1,\dots1)}(\tau)=\text{sign}(\tau)$.
\end{itemize}
However, this criterion can also be applied to higher-dimensional irreps, as we will discuss later on.

Before that, let us show that the combinations of input states $\ket{\bm{i}}$ and scattering unitaries $U$ leading to suppressed transitions described in \cite{dittel_totally_2018,dittel_totally_2018a} satisfy Eq.~\eqref{eq:supprsym}, such that the bosonic and fermionic suppression laws stated above apply. 
The input state is chosen such that a given permutation $\tau\in S_N$ of the particles has the same effect as a permutation $\Pi$ of the modes $\{\ket{0},\ket{1},\dots\ket{M-1}\}$: there exist $\tau\in S_N$ and a $M\times M$ permutation matrix $\Pi$ such that 
	\begin{align}\label{invarin}
		&\hat{R}(\tau) \ket{\bm{i}}=	\Pi^{\otimes N} \ket{\bm{i}},\\
		& \quad \text{i.e.,}\quad
		\forall \alpha \in\{1,\dots N\}, \qquad \Pi\ket{i_\alpha}=\ket{i_{\tau^{-1}(\alpha)}}.\notag
	\end{align}
 The unitary transformation $U$ is then chosen such that it diagonalizes $\Pi$, i.e., $U \Pi U^\dagger = D$, with $D$ a diagonal unitary: $D_{l,m}=\delta_{l,m} \e^{\i\phi_m}$.
For all $\sigma\in S_N$, we then have
\begin{align}
\braket{\bm{o}|\hat{R}(\sigma)^\dagger U^{\otimes N}\hat{R}(\tau)|\bm{i}}
	&=\braket{\bm{o}|\hat{R}(\sigma)^\dagger U^{\otimes N}\Pi^{\otimes N}|\bm{i}}\notag \\
	&=	\braket{\bm{o}|\hat{R}(\sigma)^\dagger D^{\otimes N} U^{\otimes N}|\bm{i}}\\
	&=\Lambda \braket{\bm{o}|\hat{R}(\sigma)^\dagger U^{\otimes N}|\bm{i}}\notag
\end{align}
with
\begin{align}
	\Lambda=\prod_{\alpha=1}^N D_{o_\alpha,o_\alpha}=\exp\left(\i \sum_{\alpha=1}^N \phi_{o_\alpha}\right).
\end{align}
Note that for given input state $\ket{\bm{i}}$ and unitary transformation $U$, $\Lambda$ depends only on the occupations of the output modes. Output configurations associated with the same value of $\Lambda$ therefore display the same suppressions.

In order to better understand which transitions can be suppressed by this mechanism beyond the bosonic and fermionic sectors, we now take a closer look at the spectrum of the operator $\hat{\rho}^{(\lambda)}(\tau)$.
For this, we restrict the representation $\lambda$ of $S_N$ to the cyclic group
\begin{align}
	\braket{\tau}=\{\tau^k|k=0, \dots p-1\},
\end{align}
where $p$ is the order of $\tau$, i.e., the smallest strictly positive integer such that $\tau^p=\id$. Since $\braket{\tau}$ is Abelian, its irreps are all one-dimensional,
so decomposing $\rho^{(\lambda)}\big|_{\braket{\tau}}$ into irreps of $\braket{\tau}$ amounts to diagonalizing $\hat{\rho}^{(\lambda)}(\tau)$ [recall Eq.~\eqref{irrepdecomp}].
Looking back to our discussion of the Fourier transform over a cyclic group in Sec.~\ref{sec:FT}, the   irreps of $\braket{\tau}$ are given by the $p$th roots of unity, $\exp(2\i\pi l/p)$, for $l=0,\dots p-1$. The matrix $\hat{\rho}^{(\lambda)}(\tau)$ therefore has the eigenvalue $\exp(2\i\pi l/p )$ with  multiplicity $m_l$ given by Eq.~\eqref{mult}:
 \begin{align}
 	m_l= \frac{1}{p}\sum_{q=0}^{p-1} \chi^{(\lambda)}(\tau^q) \exp(2\i\pi l q/p).
 \end{align}
The symmetry-induced suppression law can thus be formulated as follows: for any irrep $\lambda$ of $S_N$,
if there exists a permutation $\tau$ of order $p$ and a complex number $\Lambda$ such that
 \begin{align}
\delta_\tau * a= \Lambda a \qquad \text{and} \qquad
 		\sum_{q=0}^{p-1} \chi^{(\lambda)}(\tau^q) \Lambda^q=0,
\end{align}
then $\hat{a}(\lambda)$ vanishes.

Beyond the bosonic and fermionic suppression laws formulated above, the case of the standard irrep $\lambda=(N-1,1)$ is of particular interest.
Its character $\chi^{(N-1,1)}(\sigma)$ is given by the number of fixed points of $\sigma$ minus one \cite{hamermesh_group_1989}. If $\tau$ is a cyclic permutation [i.e., it consists of a single cycle of length $N$, e.g., $\tau=(1\, 2\dots N)$], then it has order $p=N$ and $\tau^q$ has no fixed points for $q=1,\dots N-1$. We therefore have
\begin{align}
	\sum_{q=0}^{N-1} \chi^{(N-1,1)}(\tau^q) = (N-1) +\sum_{q=1}^{N-1} (-1)=0,
\end{align}
where we have used the fact that $\tau^0=\id$ has $N$ fixed points.
We conclude that if, \emph{for a cyclic permutation} $\tau$, the transition amplitude function satisfies $\delta_\tau*a=\Lambda a$ with $\Lambda=1$, such that the bosonic transition is allowed, then  the transition is suppressed in the sector $\lambda=(N-1,1)$.
In other words, \textit{transitions that are allowed for bosons are suppressed in sector $\lambda=(N-1,1)$, and vice versa}.
A similar complementarity exists between the irreps $\lambda=(1,1,\dots)$ and $\lambda=(2,1,\dots)$, which are obtained by conjugation from the bosonic and standard irreps, respectively (recall Sec.~\ref{sec:irreps}): if $\tau$ is cyclic, transitions that are allowed for fermions are suppressed in sector $(2,1,\dots)$, and vice versa.

For the moment, we have only considered applying $\delta_\tau$ to the left of $a$, which  corresponds to performing the permutation $\tau$  on the input state. Analogously, we can consider transition amplitude functions which transform as $a*\delta_\tau=\Lambda a$ under a permutation $\tau$ of the output state.
Then $\hat{a}(\lambda)\hat{\rho}^{(\lambda)}(\tau)=\Lambda\hat{a}(\lambda)$, 
or, taking the transpose 
$\hat{\rho}^{(\lambda)}(\tau)^\top\hat{a}(\lambda)^\top=\Lambda\hat{a}(\lambda)^\top$,
so $\hat{a}(\lambda)$ vanishes if $\hat{\rho}^{(\lambda)}(\tau)^\top$ does not admit the eigenvalue $\Lambda$. Since the spectra of $\hat{\rho}^{(\lambda)}(\tau)^\top$ and $\hat{\rho}^{(\lambda)}(\tau)$ are identical, we again find that $\hat{a}(\lambda)=0$ if $\Lambda$ is not in the spectrum of $\hat{\rho}^{(\lambda)}(\tau)$.

We can further generalize by considering joint permutations  $\tau$ and $\tau'\in S_N$ of the input and output states, respectively, 
such that $\delta_\tau*a*\delta_{\tau'}= \Lambda a$. Then $\hat{\rho}^{(\lambda)}(\tau)\hat{a}(\lambda)\hat{\rho}^{(\lambda)}(\tau')=\Lambda\hat{a}(\lambda)$. This can be viewed as an eigenvalue equation for the superoperator $\hat{X}\mapsto \hat{\rho}^{(\lambda)}(\tau) \hat{X} \hat{\rho}^{(\lambda)}(\tau')$, which can be represented as the tensor product $\hat{\rho}^{(\lambda)}(\tau) \otimes \hat{\rho}^{(\lambda)}(\tau')^\top=\hat{\rho}^{(\lambda)}(\tau) \otimes \hat{\rho}^{(\lambda)}(\tau'^{-1})$. Its eigenvalues are therefore of the form $\Lambda_\tau \Lambda^{-1}_{\tau'}$, where $\Lambda_\tau$ is an eigenvalue of $\hat{\rho}^{(\lambda)}(\tau)$. If $\Lambda$ cannot be written as such a product, the transition is suppressed. We will give an example of such a suppression in Sec.~\ref{sec:fourierint}.

\subsection{Pauli-like suppressions}
\label{sec:suppr-Pauli}

Another possible reason for a component  $\hat{a}(\lambda)$ of the transition amplitude function's Fourier transform to vanish is the rule stated at the end of Sec.~\ref{sec:coset}: $\hat{a}(\lambda)=0$ whenever
the transition amplitude function $a$ hides a subgroup $H$  of $S_N$  of which $\lambda$ is a missing harmonic.
Specifically, if there exists a subgroup $H$ of $S_N$ such that
\begin{gather}\label{eq:paulisuppr1}
	\forall \tau \in H,\quad \delta_\tau * a= a\quad \text{or} 
	\quad 	\forall \tau \in H,\quad a*\delta_\tau = a,\\
	\intertext{and}
	\qquad \sum_{\tau\in H}\chi^{(\lambda)}(\tau)=0,\label{eq:paulisuppr2}
\end{gather}
then the transition is suppressed in sector $\lambda$.

In Sec.~\ref{sec:Pauli}, we have seen that the stabilizer subgroups $\stab(\bm{i})$ and $\stab(\bm{o})$ satisfy \eqref{eq:paulisuppr1}, enforcing the generalized Pauli principle.
However, for a given combination of $\ket{\bm{i}}$, $\ket{\bm{o}}$ and $U$, 
the transition amplitude function $a$ may hide another, larger subgroup $H$.
In this case, the transition can be suppressed in more sectors than predicted by the generalized Pauli principle and we speak of \textit{Pauli-like suppressions}.	
In particular, if $H$ is a Young subgroup, we can use Gamas' theorem to determine which are its missing harmonics (recall Sec.~\ref{sec:Pauli}).



There are similarities between Pauli-like suppressions and the symmetry-induced suppressions discussed in the previous section:
In both cases, the suppression is due to the absence of a one-dimensional irrep of $H$ from the restriction of an irrep $\lambda$ of $S_N$ to $H$.
For symmetry-induced transitions, the subgroup in question is the cyclic subgroup $H=\braket{\tau}$ generated by a  permutation $\tau$ of order $p$, which is Abelian, and the missing one-dimensional irrep is associated with a $p$th root of unity $\Lambda$. 
In the case of Pauli-like suppressions, the subgroup $H$ can be non-Abelian (typically a Young subgroup) and the missing one-dimensional irrep is the trivial one.
In some cases, the two types of suppression mechanisms coincide:  
  for example, the symmetry-induced suppressions in the sector $\lambda=(N-1,1)$, in the case where $\tau$ is a cyclic permutation and $\Lambda=1$ in Eq.~\eqref{eq:supprsym}, also fall under our definition of a Pauli-like suppression, with $H=\braket{\tau}$ in Eqs.~\eqref{eq:paulisuppr1} and \eqref{eq:paulisuppr2}.

Note that if only the first condition in Eq.~\eqref{eq:paulisuppr1} is satisfied, we can use Eqs.~\eqref{lcosetdecomp} and \eqref{indicator} to write
\begin{align}
	\hat{a}(\lambda)&=	\left( 	\sum_{\sigma\in S_N/H} a(\sigma)  \hat{\rho}^{(\lambda)}(\sigma)\right)  |H| \hat{I}_H(\lambda), 
\end{align}
and although $\hat{I}_H(\lambda)\neq 0$, it still may happen that the first sum (with $N!/|H|$ terms) vanishes. 
For example,  for one-dimensional irreps and small enough $S_N/H$, it is a scalar sum which can potentially be worked out explicitly, as we will see in Sec.~\ref{sec:fourierint}. This goes to show that symmetry-induced and Pauli-like suppressions do not exhaust all instances of completely destructive interference (see also \cite{bezerra_families_2023} for a description of suppressions in non-symmetric interferometers). Whether additional suppression laws rooted in symmetries can be formulated, and whether there are overarching symmetry principles unifying the known suppression mechanisms, remain open questions.

\subsection{The Fourier interferometer}
\label{sec:fourierint}

\begin{figure*}
	\begin{tabular}{c c c c c c}
		$\lambda=(4)$ & 	$\lambda=(3,1)$ & 	$\lambda=(2,2)$\\
		\includegraphics[width=0.3\textwidth]{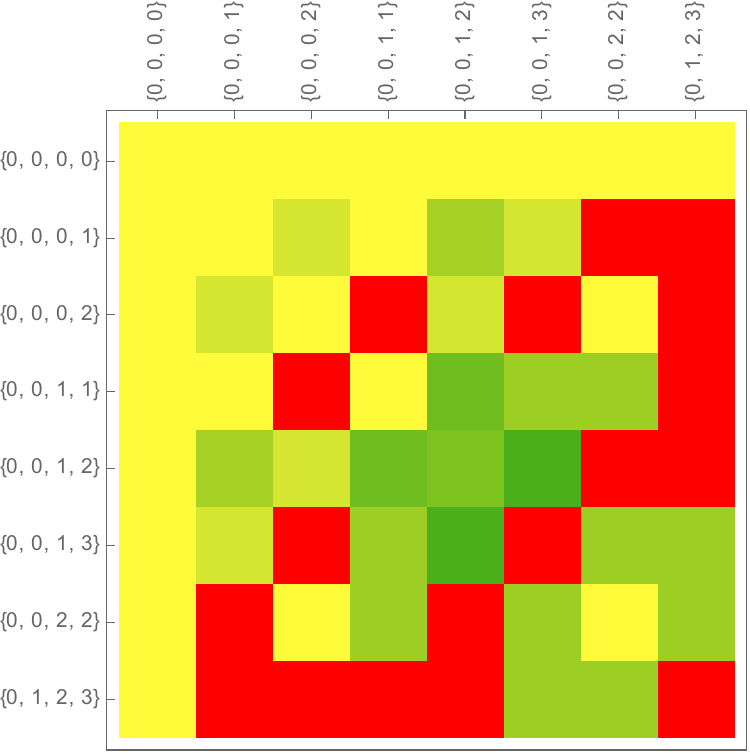}&
		\includegraphics[width=0.3\textwidth]{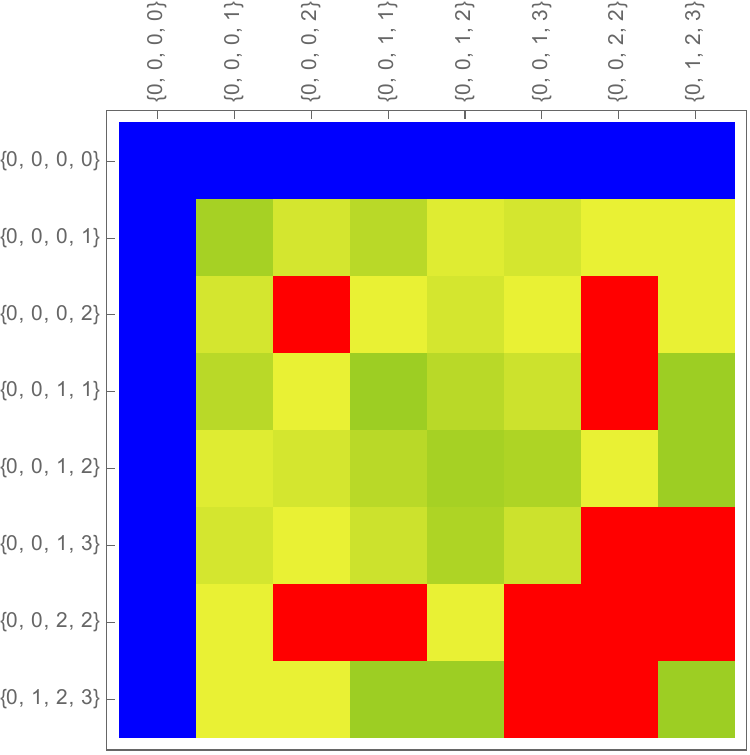}&
		\includegraphics[width=0.3\textwidth]{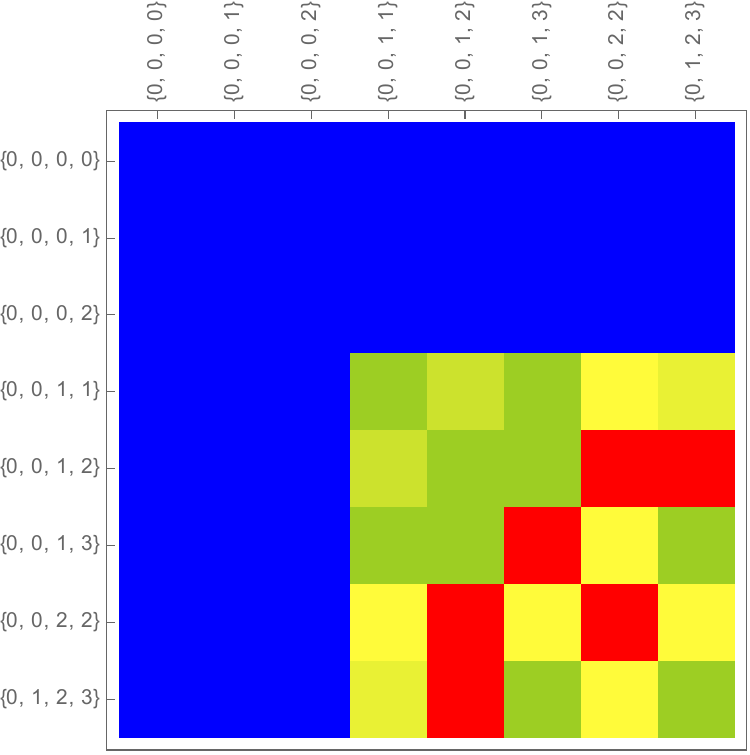}\\[2ex]
		$\lambda=(2,1,1)$ & 	$\lambda=(1,1,1,1)$ &  $\Tr\left[ \hat{a}(\lambda)^\dagger\hat{a}(\lambda)\right]$\\
		\includegraphics[width=0.3\textwidth]{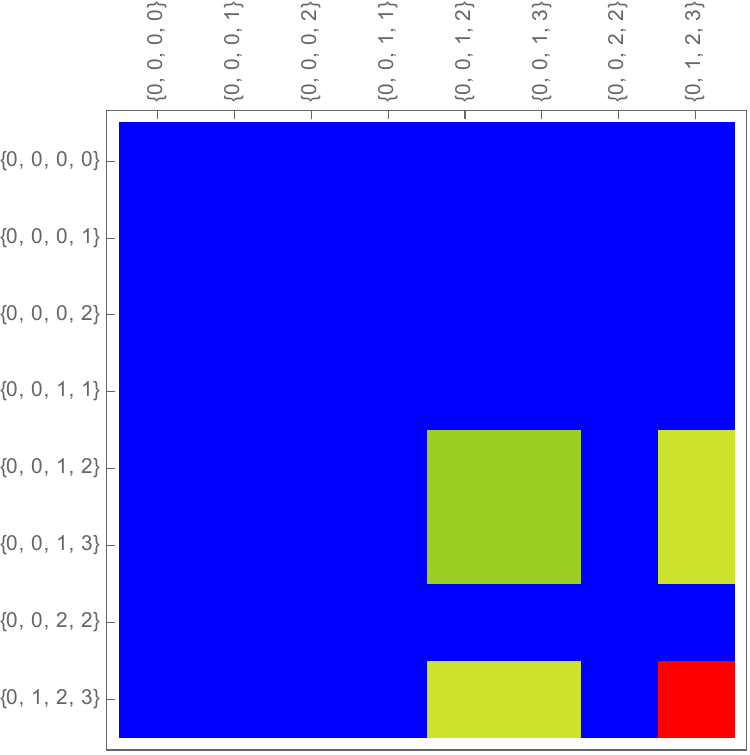}&
		\includegraphics[width=0.3\textwidth]{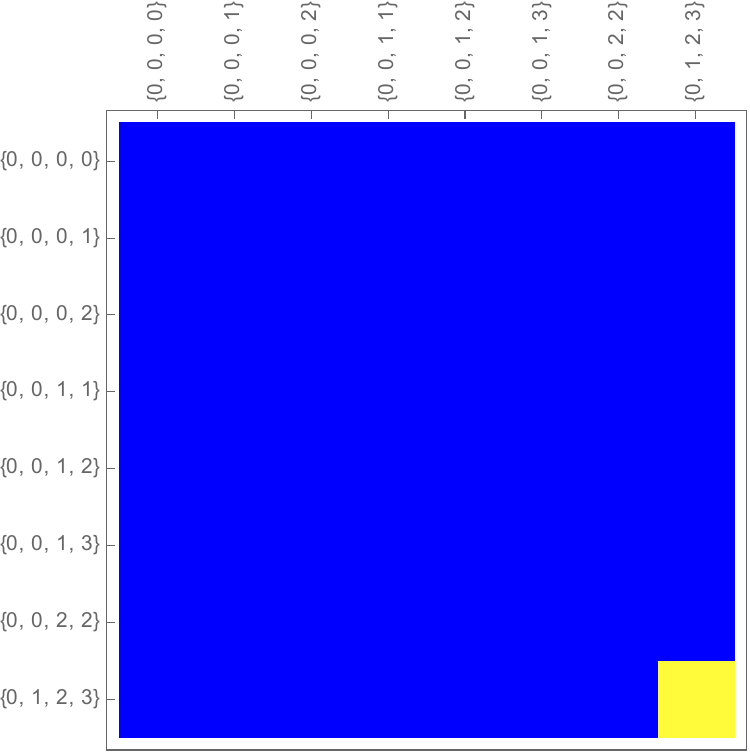}& 
		\includegraphics[height=4.8cm]{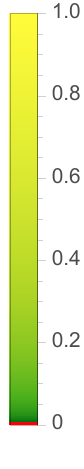}
	\end{tabular}
	
	\caption{\label{FourierN4M4}Spectral density [Eq.~\eqref{spectdens}] of the transition amplitude function [Eq.~\eqref{transamp}] for $N=4$ particles in a Fourier interferometer [Eq.~\eqref{UFT}] with $M=4$ modes. The colour code gives $\Tr\left[ \hat{a}(\lambda)^\dagger\hat{a}(\lambda)\right]$ for (inequivalent) pairs $\ket{\bm{i}}$ and $\ket{\bm{o}}$ of input and output states (labels along the x- and y-axes, respectively), for each symmetry type $\lambda$. Transitions to or from Pauli-forbidden  states (see Sec.~\ref{sec:Pauli}) are marked in blue. Otherwise suppressed transitions are marked in red.}

\end{figure*}

\begin{figure*}[p]

	\includegraphics[width=0.45\textwidth]{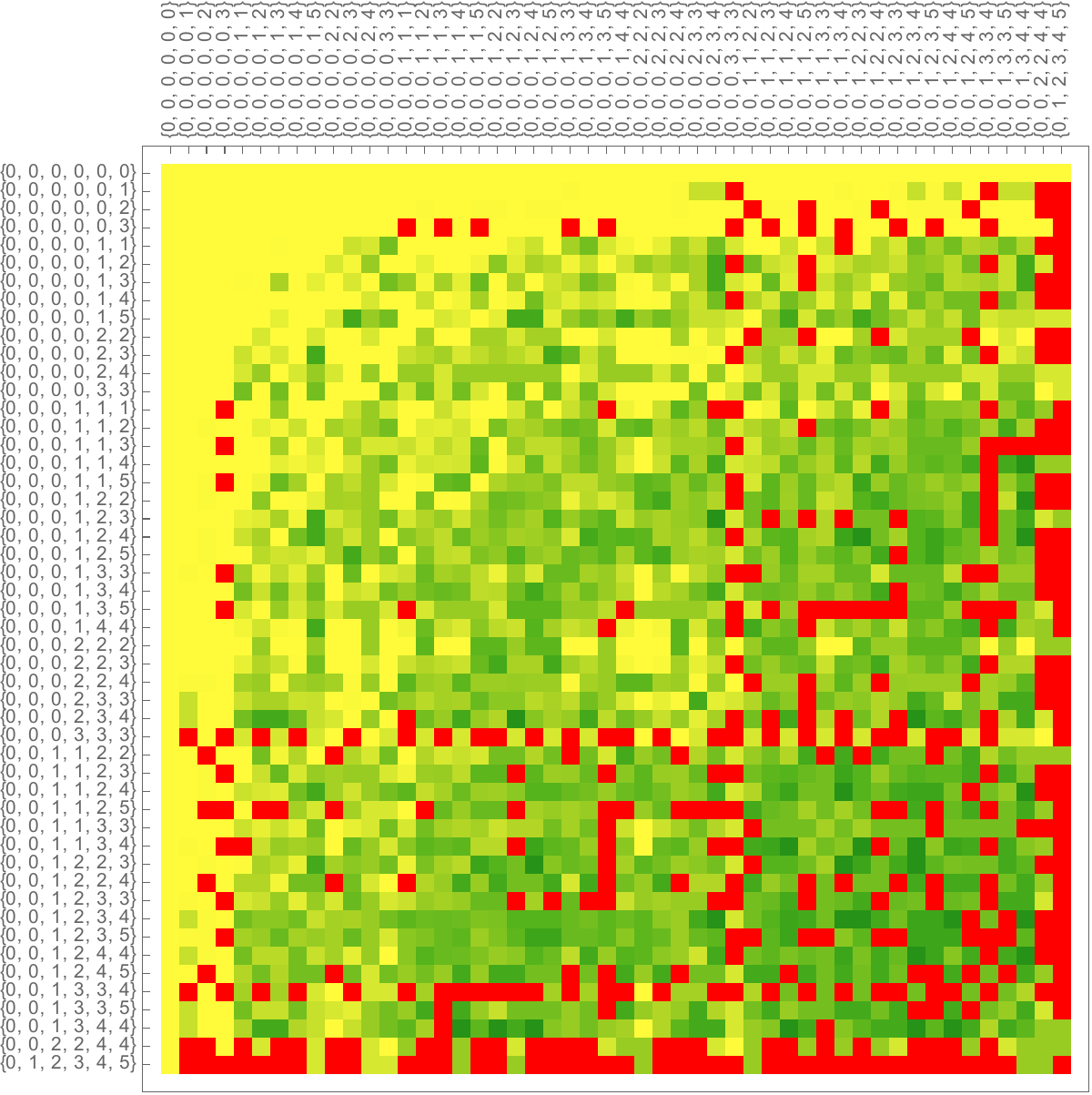}
		\includegraphics[height=6cm]{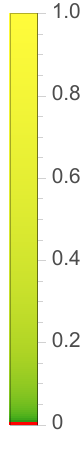}
		
		\vspace*{-6.75cm}
		\hspace{9.5cm}$\Tr\left[ \hat{a}(\lambda)^\dagger\hat{a}(\lambda)\right]$
		\vspace*{6.75cm}
	
	\includegraphics[width=0.45\textwidth,trim={0 0 0 2.6cm},clip]{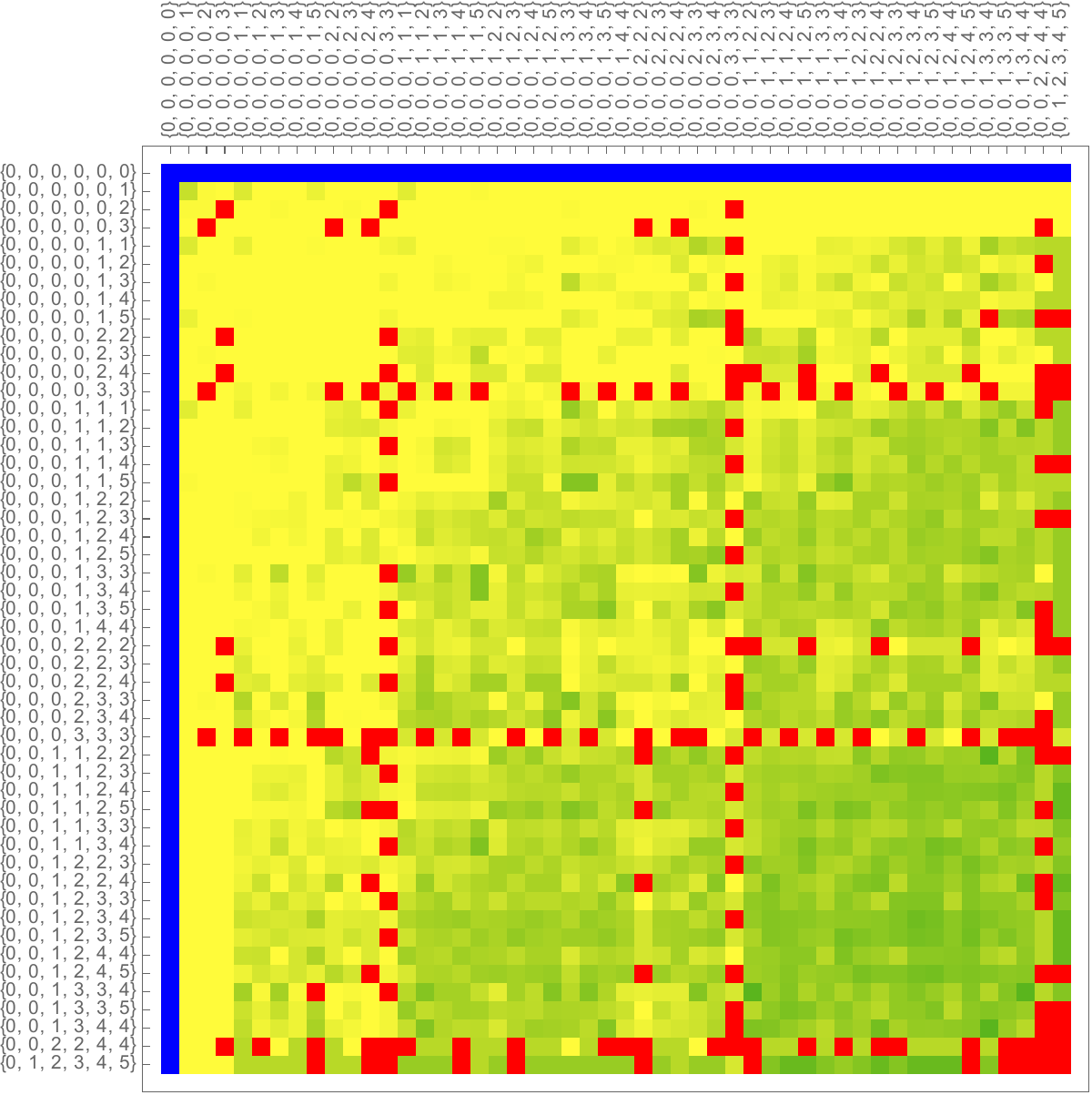} 
		\includegraphics[height=6cm]{Figures/cb6.pdf}
	
	\includegraphics[width=0.45\textwidth,trim={0 0 0 2.6cm},clip]{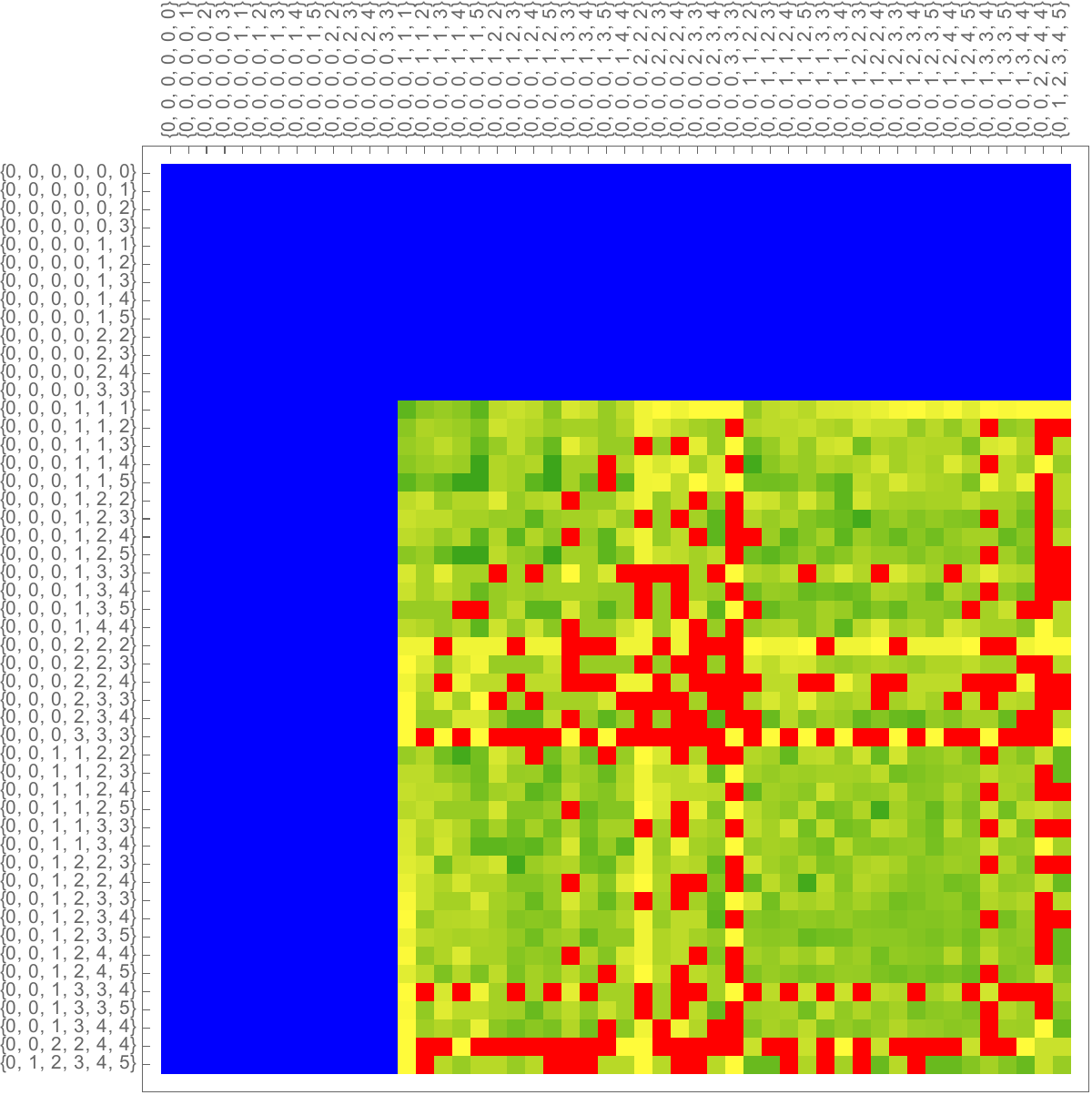}
		\includegraphics[height=6cm]{Figures/cb6.pdf}
	
	\caption{\label{FourierN6M6}Spectral density [Eq.~\eqref{spectdens}] of the transition amplitude function [Eq.~\eqref{transamp}] for $N=6$ particles in a Fourier interferometer [Eq.~\eqref{UFT}] with $M=6$ modes. The colour code gives $\Tr\left[ \hat{a}(\lambda)^\dagger\hat{a}(\lambda)\right]$ for (inequivalent) pairs $\ket{\bm{i}}$ and $\ket{\bm{o}}$ of input and output states (labels along the x- and y-axes, respectively), for  $\lambda=(6)$ (top), $\lambda=(5,1)$ (middle) and $\lambda=(3,3)$ (bottom). Transitions to or from Pauli-forbidden  states (see Sec.~\ref{sec:Pauli}) are marked in blue. Otherwise suppressed transitions are marked in red.}
	
\end{figure*}

We now illustrate the suppression mechanisms described in the previous sections by taking the example of the Fourier interferometer.
The single-particle unitary $U$ describing linear scattering through this interferometer is the $M$-dimensional discrete Fourier transform defined by
\begin{align}\label{UFT}
	\braket{l|U|m}=\frac{1}{\sqrt{M}} \exp\left( \frac{2\i\pi l m}{M}   \right),
\end{align}
 $l,m=0,\dots M-1$.
We are therefore dealing here with a Fourier transform over the cyclic group of order $M$ [Eq.~\eqref{DFT}] applied to the complex components of $M$-dimensional vectors (the single-particle states). It should not be confused with the Fourier transform over the symmetric group $S_N$.
The many-body evolution performed by the Fourier unitary is known to display symmetry-induced suppressions for bosonic and fermionic states, as discussed in \cite{lim_generalized_2005,tichy_zerotransmission_2010,tichy_manyparticle_2012,tichy_stringent_2014,dittel_totally_2018,dittel_totally_2018a}. As we will now show, symmetry-induced suppressions also occur in other symmetry sectors. In addition, we will see that the Fourier interferometer also displays Pauli-like suppressions.
As a first illustration, in Figs.~\ref{FourierN4M4} and \ref{FourierN6M6}, red boxes indicate suppressed transitions in a given symmetry sector $\lambda$ for $N=M=4$ and $N=M=6$, respectively.

The many-body transition amplitude \eqref{transamp} for the Fourier unitary \eqref{UFT},  input state $\ket{\bm{i}}$  and output state $\ket{\bm{o}}$ reads
\begin{align}\label{eq:aF}
	a(\sigma)=M^{-N/2} \exp\left(\frac{2\i \pi}{M}  \sum_{\alpha=1}^N o_\alpha i_{\sigma(\alpha)} \right),
\end{align}
for $\sigma\in S_N$, and it takes values (up to the prefactor $M^{-N/2}$) in the set of $M$th roots of unity.
It is therefore convenient to define 
\begin{align}\label{eq:phi}
	\phi(\sigma) &=  \frac{M}{2\i \pi} \log M^{N/2} a(\sigma)\notag \\ &= \sum_{\alpha=1}^N o_\alpha i_{\sigma(\alpha)} \mod M,
\end{align}
which encodes the phase of $a(\sigma)$.
We now look into how the above functions transform under permutations of the modes. This is not only relevant to identify symmetry-induced suppressions (recall Sec.~\ref{sec:suppr-sym}), but it will also allow us to identify \textit{equivalent} pairs of initial and final states, for which $\Tr\left[ \hat{a}(\lambda)^\dagger \hat{a}(\lambda)\right]$ remains unchanged, such that we can reduce the number of distinct transitions  that need to be considered.
In the following, we label the modes modulo $M$, such that
the $m$th mode, with $0\leq m\leq M-1$, is labelled by any integer of the form $m+k M$, $k\in\mathbb{Z}$. 

The cyclic group of order $M$ acts on the set of modes by translations $\Pi_q: \ket{m} \mapsto \ket{m+q},\, q=0,1,\dots M-1$, all of which are diagonalized by the Fourier unitary $U$.
Applying $\Pi_q$ to the input state $\ket{\bm{i}}$ amounts to replacing 
$i_\alpha$ by $i_\alpha+q$ in Eq.~\eqref{eq:phi}, such that 
$\phi(\sigma)$ receives an additional contribution $q\sum_{\alpha=1}^N o_\alpha \mod M$, corresponding to a rotation of $a(\sigma)$ in the complex plane.
The cyclic group can be complemented by reflections $\Sigma_q:\ket{m}\mapsto\ket{q-m},\, q=0,1,\dots M-1,$ (obeying $\Sigma_q=\Sigma_q^{-1}=\Sigma_q^\dagger$) to generate the dihedral group $D_M$: the symmetry group of a regular polygon whose vertices are identified with the modes.
Applying $\Sigma_q$ to the input state changes $\phi(\sigma)$ to $\left( q\sum_{\alpha=1}^N o_\alpha-\phi(\sigma)\right)  \mod M$, amounting to a combined complex conjugation (i.e., reflection across the real axis) and rotation of $a(\sigma)$.  
Applying an operation of the dihedral group $D_M$ to the input modes 
therefore results in a (in general different) dihedral transformation of the distribution of many-body transition amplitudes in the complex plane. The same holds for dihedral transformations of the output modes.

Since the irrep matrices $\hat{\rho}^{(\lambda)}(\sigma)$ are chosen to be real,
complex conjugation and/or multiplication by a phase of the transition amplitudes $a(\sigma)$ amount to complex conjugation and/or multiplication by a phase of the matrices $\hat{a}(\lambda)$, which leaves $\Tr\left[ \hat{a}(\lambda)^\dagger \hat{a}(\lambda)\right]$ unchanged. The dihedral transformations of input and/or output modes described above therefore all leave  $\Tr\left[ \hat{a}(\lambda)^\dagger \hat{a}(\lambda)\right]$ unchanged. In addition, applying an arbitrary permutation $\hat{R}(\sigma)$ to $\ket{\bm{i}}$ or $\ket{\bm{o}}$ results in a shift of the argument of the transition amplitude function $a$ [recall Eq.~\eqref{shift}], such that its Fourier transform $\hat{a}(\lambda)$ is multiplied by a unitary matrix $\hat{\rho}^{(\lambda)}(\sigma)$ [recall Eq.~\eqref{actSn}]. This operation therefore also leaves $\Tr\left[ \hat{a}(\lambda)^\dagger \hat{a}(\lambda)\right]$ invariant. 
Finally, exchanging the input and output states in Eq.~\eqref{eq:aF} maps $a(\sigma)$ to $a(\sigma^{-1})$ and---given that $\hat{\rho}^{(\lambda)}(\sigma^{-1})=\hat{\rho}^{(\lambda)}(\sigma)^\top$--- it maps $\hat{a}(\lambda)$ to $\hat{a}(\lambda)^\top$. The spectral density $\Tr\left[ \hat{a}(\lambda)^\dagger \hat{a}(\lambda)\right]$ is again unchanged. These invariance properties allow to drastically reduce the number of pairs of input and output states that need to be considered when searching for suppressed transitions.

Symmetry-induced suppressions, as defined in Sec.~\ref{sec:suppr-sym}, occur when the distribution of the particles over the input and/or output modes is invariant under one of the 
operations of the dihedral group described above: a common translation $\Pi_q$ or a mirroring $\Sigma_q$ of the modes for all particles.
In other word, up to a rearrangement of the particles by a permutation $\tau$, the input state $\ket{\bm{i}}$ (and/or output state $\ket{\bm{o}}$) is invariant under such a transformation.
From the above discussion, we know that these symmetries of the states translate into symmetries of the distribution of transition amplitudes $a(\sigma)$ in the complex plane.
We identify three symmetries leading to suppressed transitions:
\begin{itemize}
	\item Translational invariance of the input state:
	\begin{align}
	\hat{R}(\tau) \ket{\bm{i}}=	\Pi_q^{\otimes N} \ket{\bm{i}},
	\end{align}
	 such that
	 \begin{align}
	 \delta_\tau * a=\Lambda a, \quad \text{with} \quad \Lambda=\exp\left( \frac{2\i\pi q}{M}\sum_{\alpha=1}^N o_\alpha \right).
	\end{align}
	
	\item Translational invariance of the output state:
	\begin{align}
		\hat{R}(\tau) \ket{\bm{o}}=	\Pi_q^{\otimes N} \ket{\bm{o}}, 
	\end{align}
		such that
	\begin{align}
 a*\delta_\tau=\Lambda a, \quad \text{with} \quad \Lambda=\exp\left( \frac{2\i\pi q}{M}\sum_{\alpha=1}^N i_\alpha \right).
	\end{align}

	\item Joint mirror symmetries of input and output states:
	\begin{align}
		\hat{R}(\tau) \ket{\bm{i}}=	\Sigma_q^{\otimes N} \ket{\bm{i}}\qquad \text{and}\qquad  \hat{R}(\tau') \ket{\bm{o}}=	\Sigma_{q'}^{\otimes N} \ket{\bm{o}},
	\end{align}
		such that
	\begin{align}
     &\qquad \delta_\tau *a*\delta_{\tau'}=\Lambda a,\\
      &\text{with}\quad  \Lambda=\exp\left[ \frac{2\i\pi}{M}  \left( Nqq'-q\sum_{\alpha=1}^N o_\alpha-q'\sum_{\alpha=1}^N i_\alpha \right)\right].\notag
	\end{align}
	
\end{itemize}

\begin{figure*}
	 $\bm{i}=(0,1,2,3,4,5)$ 
	
	\includegraphics[width=0.75\linewidth]{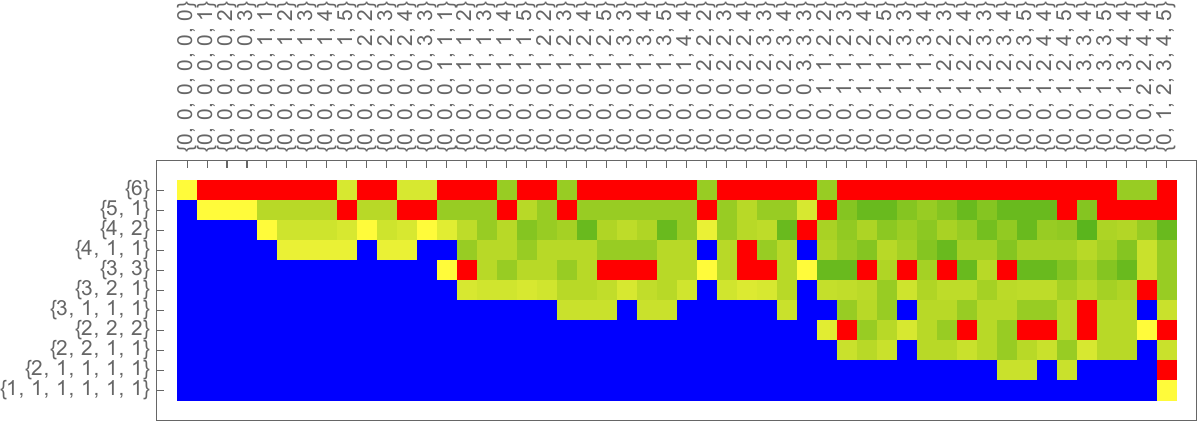}
		\includegraphics[height=4.5cm]{Figures/cb6.pdf}
		
		\vspace*{-5cm}
		\hspace{14.5cm}$\Tr\left[ \hat{a}(\lambda)^\dagger\hat{a}(\lambda)\right]$
		\vspace*{5cm}

	$\bm{i}=(0,0,2,2,4,4)$
	
	\includegraphics[width=0.75\linewidth]{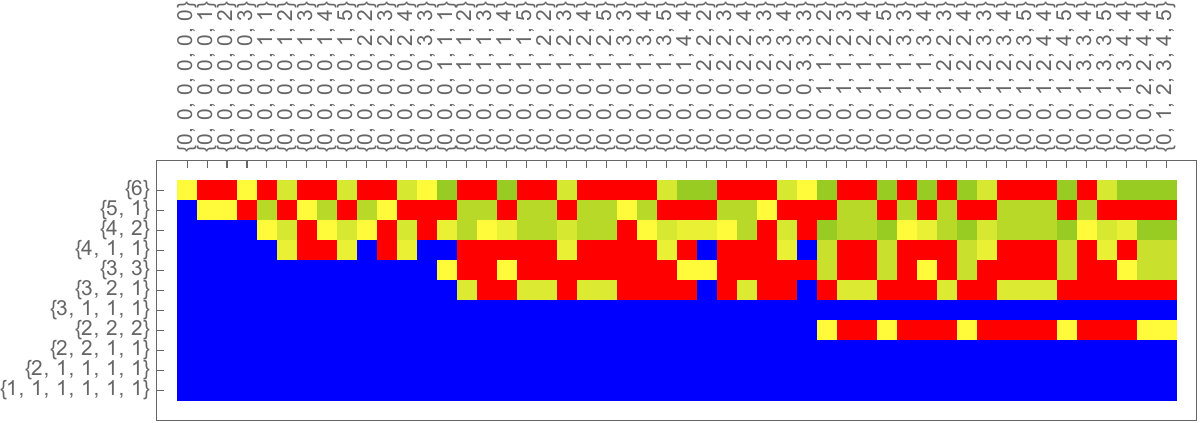}
		\includegraphics[height=4.5cm]{Figures/cb6.pdf}
	
	 $\bm{i}=(0,0,1,3,3,4)$
	 
	\includegraphics[width=0.75\linewidth]{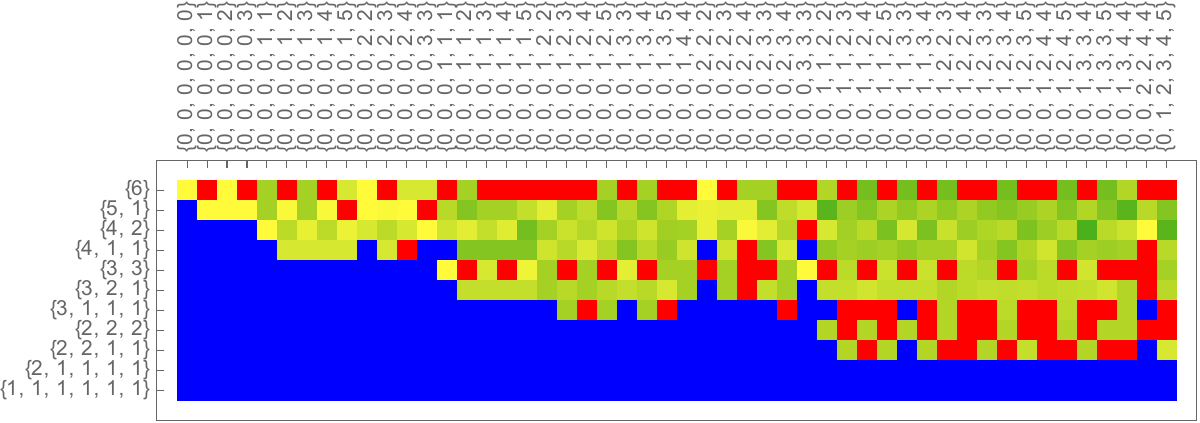}
		\includegraphics[height=4.5cm]{Figures/cb6.pdf}
	
	\caption{\label{symsuppr} Spectral density [Eq.~\eqref{spectdens}] of the transition amplitude function [Eq.~\eqref{transamp}] for three input states $\ket{\bm{i}}$ with periodic occupations in a Fourier interferometer [Eq.~\eqref{UFT}] with $M=N=6$. The colour code 
	gives the value of $\Tr\left[ \hat{a}(\lambda)^\dagger\hat{a}(\lambda)\right]$ for 
 (inequivalent) outputs	$\bm{o}$ (x-axis labels), and irreps $\lambda$ (y-axis labels). 
   Transitions to or from Pauli-forbidden  states (see Sec.~\ref{sec:Pauli}) are marked in blue. Suppressed transitions are marked in red.}
\end{figure*}

To illustrate suppressions due to translational invariance, we consider transitions from the following three input states with periodic occupations for a system of $N=6$ particles in $M=6$ modes:
\begin{itemize}
\item $\bm{i}=(0,1,2,3,4,5)$, such that $\hat{R}(\tau) \ket{\bm{i}}=	\Pi_q^{\otimes N} \ket{\bm{i}}$
with $q=1$ and $\tau=(1\, 2\, 3\, 4\, 5\, 6)$.

\item $\bm{i}=(0,0,2,2,4,4)$, such that $\hat{R}(\tau)
\ket{\bm{i}}=	\Pi_q^{\otimes N} \ket{\bm{i}}$
with $q=2$ and $\tau=(1\, 3\, 6\, 2\, 4\, 5)$.

\item $\bm{i}=(0,0,1,3,3,4)$, such that $\hat{R}(\tau)
\ket{\bm{i}}=	\Pi_q^{\otimes N} \ket{\bm{i}}$
with $q=3$ and $\tau=(1\, 4\, 2\, 5)(3\, 6)$. Note that in this case,  $\tau$ is not a cyclic permutation.

\end{itemize} 
The corresponding values of $\Tr\left[ \hat{a}(\lambda)^\dagger\hat{a}(\lambda)\right]$ for all inequivalent output states $\ket{\bm{o}}$ are shown in Fig.~\ref{symsuppr}, with suppressed transitions marked in red. Notice the complementarity of suppressions between the bosonic and standard sectors in the first two cases, where a cyclic permutation $\tau$ exists. However, it can happen that both bosonic and standard transition amplitudes vanish, although one of them is in principle allowed, for example in the survival probability of the state with one particle per mode (last column in the top panel of Fig.~\ref{symsuppr}). In this particular scenario, the suppressions are ``inherited'' from the $N=5$ case: for 5 indistinguishable bosons in distinct input modes of the $M=6$ Fourier interferometer, there are no allowed transitions to states with exactly one unoccupied output. Adding a sixth particle, whether it is distinguishable from the others or not, simply multiplies the --- vanishing --- 5-particle transition amplitudes by an extra factor.

\begin{figure*}
	
	\includegraphics[width=0.25\linewidth]{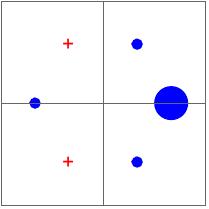}
	\includegraphics[width=0.25\linewidth]{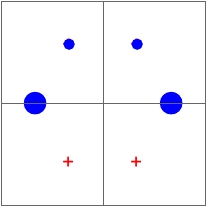}	
    \includegraphics[width=0.25\linewidth]{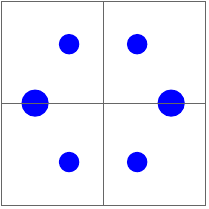}	

	\caption{\label{000135_001233}Left: Distribution of $N=6$ particles over $M=6$ modes, identified with the vertices of a hexagon, for the input mode list $\bm{i}= (0, 0, 0, 1, 3, 5)$. Notice the mirror symmetry across the x-axis. Middle: same for the output mode list $\bm{o}= (0, 0, 1, 2, 3, 3)$. Notice the symmetry across the y-axis. Right: Corresponding distribution in the complex plane of the many-body transition amplitudes $a(\sigma)=\braket{\bm{o}|R(\sigma)^\dagger U^{\otimes N}|\bm{i}}$, for $\sigma$ running over $S_N$, when $U$ is the Fourier unitary \eqref{UFT}. Notice the joint mirror symmetry across the x- and y-axes, resulting in a symmetry around the origin. The radius of the blue dots is proportional to the number of particles or amplitudes, red crosses indicate unoccupied modes or absent amplitudes.}
\end{figure*}

As an example of a suppression due to joint  mirror symmetries of input and output, consider again $N=M=6$. The input state with $\bm{i}= (0, 0, 0, 1, 3, 5)$ satisfies $\hat{R}(\tau) \ket{\bm{i}}=	\Sigma_q^{\otimes N} \ket{\bm{i}}$ with $q=0$ and $\tau=(4\, 6)$. The output	state with	
$\bm{o}= (0, 0, 1, 2, 3, 3)$ satisfies $\hat{R}(\tau') \ket{\bm{o}}=	\Sigma_{q'}^{\otimes N} \ket{\bm{o}}$ with 	
	$q'=3$ and $\tau'= (1\, 6)(2\, 5)(3\, 4)$. We therefore have  $\delta_\tau *a*\delta_{\tau'}=\Lambda a$ with $\Lambda=-1$. 
	Indeed, as shown in Fig.~\ref{000135_001233}, the distribution of many-body transition amplitudes is symmetric around the origin of the complex plane, leading to a suppression of the bosonic transition.  

Pauli-like suppressions arise from the invariance of the transition amplitude function under the action of a subgroup of $S_N$. In the extreme case of a constant function $a(\sigma)=a$, it is clear that $\hat{a}(\lambda)$ must vanish for all but the bosonic irrep. This can for example happen when particles only occupy equally separated input and output modes. Specifically, if particles on
input only occupy modes $i$ that are a multiple of $p_i\in \mathbb{Z}$, while particles on output only occupy modes $o$ that are a multiple of $p_o\in \mathbb{Z}$, and if these periods satisfy $p_ip_o \mod M = 0$, then it follows from Eq.~\eqref{eq:phi} that, for all $\sigma$, $\phi(\sigma)=0$, such that $a(\sigma)$ is constant.

\begin{figure*}

	\includegraphics[width=0.25\linewidth]{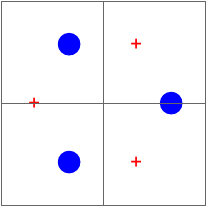}
	\includegraphics[width=0.25\linewidth]{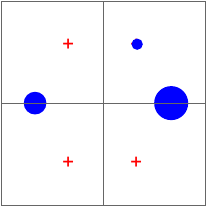}	
	\includegraphics[width=0.25\linewidth]{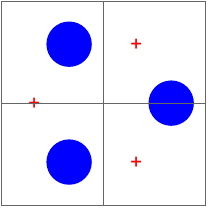}	
	
	\caption{\label{002244_000133}Left: Distribution of $N=6$ particles over $M=6$ modes, identified with the vertices of a hexagon, for the input mode list $\bm{i}= (0, 0, 2, 2, 4, 4)$. Occupied modes are all multiples of $p_i=2$.  Middle: same for the output mode list $\bm{o}= (0, 0, 0, 1,  3, 3)$. A single particle occupies a mode ($o=1$) which is not a multiple of $p_o=3$. As a consequence,  for the Fourier unitary \eqref{UFT}, there is a Pauli-like suppression in all sectors apart from the bosonic and standard ones.
	Because of the input state's periodicity, the bosonic transition is symmetry-suppressed, such that only the standard sector is not suppressed.
		 Right: Corresponding distribution in the complex plane of the many-body transition amplitudes $a(\sigma)=\braket{\bm{o}|R(\sigma)^\dagger U^{\otimes N}|\bm{i}}$, for $\sigma$ running over $S_N$. The three-fold rotation symmetry reflects that of the input state. The radius of the blue dots is proportional to the number of particles or amplitudes, red crosses indicate unoccupied modes or absent amplitudes.}
\end{figure*}

Now if a single output particle, say the $N$th one, populates a mode $o$ which is \emph{not} a multiple of $p_o$, then $\phi(\sigma)=o i_{\sigma(N)} \mod M$ depends only on $\sigma(N)$, and is therefore invariant under transformations of $S_{N-1}$. It follows that the transition is suppressed in all but the bosonic and standard sectors.
Note that, in this case, since $\sigma(N)$ takes each of the values $1, \dots N$ equally many times when $\sigma$ runs over $S_N$, the bosonic transition amplitude takes the simple form 
\begin{align}
 \hat{a}(N)&=M^{-N/2} (N-1)!	\sum_{\alpha=1}^N \exp\left( \frac{2\i \pi}{M} o i_\alpha  \right)\notag \\
 &=M^{-N/2} (N-1)!\sum_{m=0}^{M-1} n_m \exp\left( \frac{2\i \pi}{M} o m  \right),
\end{align}
where $n_m$ is the initial occupation of mode $m$. We recognise the $o$th component of the discrete Fourier transform of the initial particle distribution. If this vanishes, then the standard irrep is the only one in which the transition is not suppressed.
Figure \ref{002244_000133} shows an example of such a transition.
 When more particles do not populate equally spaced sites, the subgroup of permutations fixing $a(\sigma)$ becomes smaller, implying  Pauli-like suppressions in fewer  symmetry sectors and  more complex expressions for the bosonic transition amplitude. For example, if $k$ output particles all populate a mode $o$ which is not a multiple of $p_o$, the bosonic transition amplitude is proportional to
 \begin{align}
 	\hat{a}(N) \propto\sum_{\alpha_1<\alpha_2\dots< \alpha_k} \exp\left( \frac{2\i \pi}{M} o \sum_{j=1}^k i_{\alpha_j}  \right), 
 \end{align}
 which, for moderate $k$, can reasonably easily be seen to vanish or not, depending on $\bm{i}$.

 \section{Conclusion}
 
We have introduced the powerful tools of Fourier analysis to the study of many-body interference in the dynamics of identical particles.
In particular, we have investigated the Fourier transforms of the many-body transition amplitude function $a$---which gathers all $N!$ transition amplitudes between the initial and final states of $N$ particles---and of the partial distinguishability function $j$---which encodes the effect of the particles' internal states.
The non-commutativity of the group of particle permutations results in a
rich structure of the Fourier transforms, with matrix-valued mixed-symmetry components alongside the single-dimensional bosonic and fermionic components.
We have argued that this decomposition into exchange symmetry sectors
provides a natural framework to understand the effect of partial distinguishability in the interference of systems of bosons or fermions.

We envision several potential applications of the present framework to further problems in the fields of many-body physics, quantum optics and quantum information. 
Since it relies only on the commutation of the evolution operator with particle permutations, and not on its factorization as a product of single-particle unitaries, our symmetry-based approach offers a promising way of  including interaction effects to many-body interference scenarios \cite{dittel_waveparticle_2021,dufour_manybody_2020,bressanini_noise_2022,spagnolo_nonlinear_2023,brunner_manybody_2023} beyond the perturbative regime \cite{brunner_signatures_2018}.
Although we haven't addressed this possibility explicitly, we also believe that the present framework is  useful to describe many-body systems which do not have an intrinsic exchange symmetry (e.g., spin chains or qubit registers), but which are submitted to symmetric evolution and measurement protocols, rendering them ``operationally indistinguishable'' \cite{yadin_thermodynamics_2023}. In this case, the exchange symmetry $\lambda$ of an entangled state is conserved, providing a means of protecting quantum information from global noise sources.
There has been intense recent activity around bosonic suppression laws in the Fourier interferometer, with proposed applications to the
efficient benchmarking of many-photon indistinguishability  \cite{novo_native_2026,sanz_exponential_2026,schadow_certification_2026},
to the generation of entangled states \cite{bhatti_generating_2023,piccolini_robust_2024a,park_entangled_2025}, to the distillation of pure and indistinguishable photons \cite{somhorst_photondistillation_2025,saied_general_2025,somhorst_belowthreshold_2026} and to quantum metrology \cite{descamps_role_2026} and computing \cite{ustun_fusion_2025} protocols. We are confident that our investigation of completely destructive interference beyond the bosonic and fermionic symmetry sectors will allow to further develop such protocols, and to devise new ones, for example for the generation of states with a specific exchange symmetry. Finally, our discussion of suppressed transitions in the Fourier interferometer has uncovered an interesting interplay between two
Fourier transforms: the one over the permutation group of the particles ($S_N$) and the other over the translation group of the modes ($\mathbb{Z}/M\mathbb{Z}$). The full consequences of this interplay are yet to be elucidated.

\clearpage

 \bibliography{groupFT}

\clearpage

\onecolumngrid

\section*{Appendices}

\appendix

\section{Fourier inversion formula}
\label{app:IFT}

We give a brief justification for the Fourier inversion formula, Eq.~\eqref{IFT}. It can be understood in light of the Grand Orthogonality Theorem:  matrix irreps  $\rho^{(\lambda)}$,  $\rho^{(\nu)}$ of a finite group $G$ satisfy \cite{hamermesh_group_1989,chen,sengupta}
\begin{align}\label{GOT}
	\sum_{\sigma\in G} \hat{\rho}_{mn}^{(\lambda)}(\sigma^{-1}) \hat{\rho}_{n'm'}^{(\mu)}(\sigma)=\frac{|G|}{d_\lambda} \delta_{\lambda\mu}\delta_{mm'}\delta_{nn'}.
\end{align}
For unitary irreps,  $\hat{\rho}_{mn}^{(\lambda)}(\sigma^{-1})=\hat{\rho}_{nm}^{(\lambda)}(\sigma)^*$, such that we can write
\begin{align}
	( \hat{\rho}_{nm}^{(\lambda)}, \hat{\rho}_{n'm'}^{(\mu)})=\frac{|G|}{d_\lambda} \delta_{\lambda\mu}\delta_{mm'}\delta_{nn'}
\end{align}
with the inner product defined in \eqref{scalarprod}.
In essence, the above relation means that 
the individual entries of the matrix irreps form an orthogonal set of functions from $G$ to $\mathbb{C}$. Since $\sum_\lambda d_\lambda^2=|G|$, this set forms a basis,  and the Fourier inversion formula Eq.~\eqref{IFT} is the expansion of  $f$ in that basis.

Setting $n=m$ and $n'=m'$ in Eq.~\eqref{GOT} and summing over $m$ and $m'$, we obtain the following relation for the irreducible characters  $\chi^{(\lambda)}(\sigma)=\Tr(\hat{\rho}^{(\lambda)}(\sigma))$:
\begin{align}
	\sum_{\sigma\in G} \chi^{(\lambda)}(\sigma^{-1}) \chi^{(\mu)}(\sigma)=|G| \delta_{\lambda\mu}.
\end{align}
Viewing the above as an orthogonality relation for the rows of the character table, we can deduce the following orthogonality relation for its columns: if $\sigma$ and $\tau$ are in the same conjugacy class $C$, then
\begin{align}
	\sum_{\lambda} \chi^{(\lambda)}(\sigma^{-1}) \chi^{(\lambda)}(\tau)
		=|G|/|C|,
\end{align}
 otherwise, the left-hand-side vanishes.
In particular, given that the group's identity element $\id$ is alone in its conjugacy class and  that $\chi^{(\lambda)}(\id)=d_\lambda$, we have 	
\begin{align}\label{charactort}
	\sum_{\lambda}\frac{d_\lambda}{|G|} \chi^{(\lambda)}(\tau)= \delta_{\tau,\id}.
\end{align}

To prove the Fourier inversion formula \eqref{IFT}, we start from the right hand side and
replace  $\hat{f}(\lambda)$ by its definition \eqref{FT}. This yields
\begin{align}
\sum_{\lambda}  \frac{d_\lambda}{|G|} \Tr \left[ \hat{\rho}^{(\lambda)}(\sigma)^\dagger \hat{f}(\lambda)\right]=\sum_{\tau\in G} f(\tau)  \sum_{\lambda} \frac{d_\lambda}{|G|} \Tr \left[ \hat{\rho}^{(\lambda)}(\sigma)^\dagger \hat{\rho}^{(\lambda)}(\tau)  \right].
\end{align}
Using the fact that $\rho^{(\lambda)}$ is a unitary representation, we arrive at a sum of characters of the form of Eq.~\eqref{charactort}, which allows us to conclude:
\begin{align}
	\sum_{\lambda}  \frac{d_\lambda}{|G|} \Tr \left[ \hat{\rho}^{(\lambda)}(\sigma)^\dagger \hat{f}(\lambda)\right]&=\sum_{\tau\in G} f(\tau)  \sum_{\lambda} \frac{d_\lambda}{|G|} \chi^{(\lambda)}(\sigma^{-1}\circ\tau)\notag \\
&=\sum_{\tau\in G} f(\tau)  \delta_{\sigma^{-1}\circ\tau,\id}\\
&=f(\sigma).\notag
\end{align}

\section{Formulas for counting statistics}

\label{app:cs}

We provide proofs of the formulas Eq.~\eqref{countstats} and Eq.~\eqref{countstat-partdist}.

 Equation \eqref{countstats} gives the counting statistics arising from an initial state $\hat{c}(R)\ket{\bm{i}}$:
\begin{align}
	\mathrm{prob}(n_0,\dots n_{M-1})&=	\frac{\braket{\bm{i}|\hat{c}(R)^\dagger U^{\dagger \otimes N} P_{\bm{o}}  U^{\otimes N} \hat{c}(R) |\bm{i}}}{\braket{\bm{i}|\hat{c}(R)^\dagger \hat{c}(R)|\bm{i}}}.
\end{align}	
By definition of the projector $P_{\bm{o}}$ [Eq.~\eqref{measurement}], the numerator reads
\begin{align}
\braket{\bm{i}|\hat{c}(R)^\dagger U^{\dagger \otimes N} P_{\bm{o}}  U^{\otimes N} \hat{c}(R) |\bm{i}}	=\frac{1}{|\stab(\bm{o})|} \sum_{\sigma\in S_N}\braket{\bm{i}|\hat{c}(R)^\dagger U^{\dagger \otimes N}  \hat{R}(\sigma) |\bm{o}}		\braket{\bm{o}| \hat{R}(\sigma)^\dagger  U^{\otimes N} \hat{c}(R) |\bm{i}}.
\end{align}	
Recognising 
\begin{align}
	\braket{\bm{o}| \hat{R}(\sigma)^\dagger  U^{\otimes N} \hat{c}(R) |\bm{i}}
	=(c*a)(\sigma),
\end{align}
we thus have
\begin{align}
	\braket{\bm{i}|\hat{c}(R)^\dagger U^{\dagger \otimes N} P_{\bm{o}}  U^{\otimes N} \hat{c}(R) |\bm{i}}
	&=\frac{1}{|\stab(\bm{o})|}  (  c*a ,c*a) 
\end{align}	
and by the Parseval-Plancherel identity \eqref{PPid},
\begin{align}
	\braket{\bm{i}|\hat{c}(R)^\dagger U^{\dagger \otimes N} P_{\bm{o}}  U^{\otimes N} \hat{c}(R) |\bm{i}}
	&=\frac{1}{|\stab(\bm{o})|} \sum_\lambda \frac{d_\lambda}{N!}\Tr\left[  \hat{a}(\lambda)^\dagger \hat{c}(\lambda)^\dagger \hat{c}(\lambda) \hat{a}(\lambda) \right].
\end{align}	
The denominator evaluates to 
\begin{align}
\braket{\bm{i}|\hat{c}(R)^\dagger \hat{c}(R)|\bm{i}}&=\sum_{\sigma,\tau\in S_N} c(\sigma)^* c(\tau) \braket{\bm{i}| \hat{R}(\sigma)^\dagger \hat{R}(\tau)|\bm{i}}\notag \\
&=|\stab(\bm{i})| \sum_{\sigma,\tau\in S_N} c(\sigma)^* c(\tau) I_{\bm{i}}(\tau^{-1}\circ\sigma)\notag\\
&= |\stab(\bm{i})| (c, c* I_{\bm{i}})\notag\\
&=|\stab(\bm{i})| \sum_\lambda \frac{d_\lambda}{N!}\Tr\left[   \hat{c}(\lambda)^\dagger \hat{c}(\lambda)\hat{I}_{\bm{i}}(\lambda) \right].
\end{align}
Altogether
\begin{align}
	\mathrm{prob}(n_0,\dots n_{M-1})	&=\frac{1}{|\stab(\bm{i})||\stab(\bm{o})|} \frac{\sum_\lambda d_\lambda\Tr\left[  \hat{a}(\lambda)^\dagger \hat{c}(\lambda)^\dagger \hat{c}(\lambda) \hat{a}(\lambda) \right]}
	{ \sum_\lambda d_\lambda\Tr\left[  \hat{c}(\lambda)^\dagger \hat{c}(\lambda) \hat{I}_{\bm{i}}(\lambda) \right]}. 
\end{align}

Equation \eqref{countstat-partdist} gives the same probability in the case of partially distinguishable particles with initial external state $\varrho$: 
\begin{align}
	\mathrm{prob}(n_0,\dots n_{M-1})=	\Tr(U^{\otimes N} \varrho U^{\dagger\otimes N} P_{\bm{o}}).
\end{align}
We assume that $\varrho$ is supported on the subspace generated by vectors of the form $\hat{R}(\sigma)\ket{\bm{i}}$ for $\sigma\in S_N$. The projector on this subspace is
\begin{align}\label{eq:Pi}
P_{\bm{i}}=	\frac{1}{|\stab(\bm{i})|}\sum_{\sigma\in S_N} \hat{R}(\sigma)\ket{\bm{i}}\bra{\bm{i}}\hat{R}(\sigma)^\dagger,
\end{align}
in analogy with the projector $P_{\bm{o}}$  of Eq.~\eqref{measurement}.
Writing $\varrho=P_{\bm{i}}\varrho P_{\bm{i}}$, we obtain
\begin{align}
	\mathrm{prob}(n_0,\dots n_{M-1})&=\frac{1}{|\stab(\bm{o})||\stab(\bm{i})|^2} \sum_{\sigma,\tau,\pi\in S_N} 
	\braket{\bm{o}| \hat{R}(\sigma)^\dagger  U^{\otimes N} \hat{R}(\tau) |\bm{i}}
	\braket{\bm{i}|\hat{R}(\tau)^\dagger \varrho \hat{R}(\pi)|\bm{i}}
	  \braket{\bm{i}|\hat{R}(\pi)^\dagger U^{\dagger \otimes N}  \hat{R}(\sigma) |\bm{o}}\notag\\
	  &=\frac{1}{N!|\stab(\bm{o})||\stab(\bm{i})|} \sum_{\sigma,\tau,\pi\in S_N}  a(\tau^{-1}\circ\sigma) j(\pi^{-1}\circ\tau) a(\pi^{-1}\circ\sigma)^*\notag\\
	  &=\frac{1}{N!|\stab(\bm{o})||\stab(\bm{i})|} \sum_{\sigma,\pi\in S_N} a(\pi^{-1}\circ\sigma)^*  (j*a)(\pi^{-1}\circ\sigma)\notag\\
	  &=\frac{1}{|\stab(\bm{o})||\stab(\bm{i})|} \sum_{\sigma\in S_N} a(\sigma)^*  (j*a)(\sigma)\notag\\
	  &=\frac{1}{|\stab(\bm{o})||\stab(\bm{i})|} (a,j*a).
\end{align}
Finally, with the Parseval-Plancherel identity,
\begin{align}
	\mathrm{prob}(n_0,\dots n_{M-1})= \frac{1}{|\stab(\bm{o})| |\stab(\bm{i})|}  \sum_\lambda \frac{d_\lambda}{N!} \Tr\left[\hat{a}(\lambda)^\dagger\hat{j}(\lambda)\hat{a}(\lambda)\right].
\end{align}

\section{Positivity of the Fourier transform of the partial distinguishability function}
\label{app:positivity}


We show that the Fourier transform $\hat{j}(\rho)$ of $j$ at any unitary representation $\rho$ of dimension $d$ is positive, i.e., for any vector $\vec{x}\in \mathbb{C}^d$, 
$\vec{x}^\dagger \hat{j}(\rho)  \vec{x}\geq 0$.

We have
\begin{align}
	\vec{x}^\dagger  \hat{j}(\rho)  \vec{x} &= \sum_{\sigma\in S_N} j(\sigma)\ \vec{x}^\dagger  \hat{\rho}(\sigma)  \vec{x}= \frac{1}{N!}\sum_{\sigma,\tau\in S_N} j(\tau^{-1}\circ\sigma)\ \vec{x}^\dagger  \hat{\rho}(\tau^{-1}\circ\sigma)  \vec{x}.
\end{align}
Given that
\begin{align}
	j(\tau^{-1}\circ\sigma)=\frac{N!}{|\stab(\bm{i})|}	\braket{\bm{i}| \hat{R}(\tau^{-1}\circ\sigma)^\dagger \varrho  |\bm{i}  }=\frac{N!}{|\stab(\bm{i})|}\braket{\bm{i}| \hat{R}(\sigma)^\dagger \varrho \hat{R}(\tau) |\bm{i}  }
\end{align}
and
\begin{align}
	\hat{\rho}(\tau^{-1}\circ\sigma)=\hat{\rho}(\tau)^\dagger\hat{\rho}(\sigma),
\end{align}
we can write
\begin{align}
	\vec{x}^\dagger  \hat{j}(\rho)  \vec{x} 
	&= \frac{1}{|\stab(\bm{i})|}\sum_{\sigma,\tau\in S_N} \braket{\bm{i}| \hat{R}(\sigma)^\dagger \varrho \hat{R}(\tau) |\bm{i}  }
	\vec{x}^\dagger  \hat{\rho}(\tau)^\dagger\hat{\rho}(\sigma)  \vec{x}\notag\\
	&=\frac{1}{|\stab(\bm{i})|}\sum_{\sigma,\tau\in S_N} \braket{\bm{i}| \hat{R}(\sigma)^\dagger \varrho \hat{R}(\tau) |\bm{i}  }	 
	\vec{y}(\tau)^\dagger  \vec{y}(\sigma).
\end{align}
where we have defined the vector valued function $\vec{y}(\sigma)=\hat{\rho}(\sigma) \vec{x}$. Decomposing it into components, we find
\begin{align}\label{positivity}
	\vec{x}^\dagger  \hat{j}(\rho)  \vec{x} 
	=\frac{1}{|\stab(\bm{i})|}\sum_{\sigma,\tau\in S_N}\sum_{k=1}^d \braket{\bm{i}| \hat{R}(\sigma)^\dagger y_k(\sigma) \varrho y_k(\tau)^*\hat{R}(\tau) |\bm{i}  }=	 \frac{1}{|\stab(\bm{i})|}\sum_{k=1}^d \braket{\psi_k | \varrho|\psi_k },
\end{align}
with 
\begin{align}
	\ket{\psi_k}= \sum_{\tau\in S_N} y_k(\tau)^*\hat{R}(\tau) \ket{\bm{i}  }.
\end{align}
By positivity of $\varrho$, every term in the right-hand side of Eq.~\eqref{positivity} is positive and we conclude that $\vec{x}^\dagger  \hat{j}(\rho)  \vec{x} \geq 0$. Finally, we note that a positive matrix is also Hermitian.

\section{Purity and sector weights  of states of partially distinguishable particles}
\label{app:weights}

We prove equations \eqref{eq:purity} and \eqref{eq:weights}.

We write the  purity of $\varrho$ as $\Tr\left[  \varrho^2 \right]=\Tr\left[  \varrho^\dagger \varrho \right]$. Given that $\varrho=P_{\bm{i}}\varrho$, with $P_{\bm{i}}$ from Eq.~\eqref{eq:Pi}, the trace evaluates to
\begin{align}
	\Tr\left[   \varrho^\dagger \varrho \right]&=\frac{1}{|\stab(\bm{i})|^2}\sum_{\sigma,\tau\in S_N} \braket{\bm{i}|\hat{R}(\sigma)^\dagger \varrho \hat{R}(\tau) |\bm{i}} \braket{\bm{i}|\hat{R}(\tau)^\dagger \varrho^\dagger    \hat{R}(\sigma)|\bm{i}},
\end{align}
which we can express in terms  of the function $j$ [Eq.~\eqref{eq:j}]:
\begin{align}
	\Tr\left[   \varrho^\dagger \varrho \right]
	&=\frac{1}{(N!)^2}\sum_{\sigma,\tau\in S_N} j(\tau^{-1}\circ\sigma) j(\tau^{-1}\circ\sigma)^*.
\end{align}
The double sum reduces to an inner product \eqref{scalarprod}, which we finally express using the Parseval-Plancherel identity \eqref{PPid} to obtain the desired result:
\begin{align}
	\Tr\left[   \varrho^\dagger \varrho \right]= \frac{1}{N!} (j,j)=\frac{1}{N!} \sum_\lambda \frac{d_\lambda}{N!} \Tr\left[ \hat{j}(\lambda)^\dagger \hat{j}(\lambda) \right].
\end{align}

Now to the proof of Eq.~\eqref{eq:weights}. The weight of a state $\varrho$ on symmetry sector $\lambda$ is given by $p_\lambda=\Tr\left[ \hat{P}^{(\lambda)}(R)  \varrho \right]=\Tr\left[  \hat{P}^{(\lambda)}(R)^\dagger \varrho \right]$. With the definition \eqref{isotyp} of the projector $\hat{P}^{(\lambda)}(R)$ and again using $\varrho=P_{\bm{i}}\varrho$, we find
\begin{align}
	p_\lambda&=\Tr\left[\hat{P}^{(\lambda)}(R)^\dagger   \varrho \right]\notag\\
	&=	\frac{1}{|\stab(\bm{i})|}\frac{d_\lambda}{N!}  \sum_{\sigma,\tau\in S_N} \chi^{(\lambda)}(\sigma) \braket{\bm{i}|\hat{R}(\tau)^\dagger \hat{R}(\sigma)^\dagger\varrho  \hat{R}(\tau)|\bm{i}}\notag\\
	&=	\frac{d_\lambda}{(N!)^2}  \sum_{\sigma,\tau\in S_N} \chi^{(\lambda)}(\sigma) j(\tau^{-1}\circ\sigma\circ\tau).
\end{align}

Changing variables from $\sigma$ to $\pi=\tau^{-1}\circ\sigma\circ\tau$, we obtain
\begin{align}
	p_\lambda
	=	\frac{d_\lambda}{(N!)^2}  \sum_{\pi,\tau\in S_N} \chi^{(\lambda)}(\tau\circ\pi\circ\tau^{-1}) j(\pi)=	\frac{d_\lambda}{N!}\sum_{\pi\in S_N} \chi^{(\lambda)}(\pi) j(\pi),
\end{align}
where we have used the fact that $\chi^{(\lambda)}$ is a class function in the second equality. Given that $\chi^{(\lambda)}$ is real, we recognize the 
inner product $(\chi^{(\lambda)},j)$, which we can express using Eq.~\eqref{PPid} and Eqs.~\eqref{hatPmu},\eqref{hatPmuchimu} to conclude that
\begin{align}
	p_\lambda	&=	\frac{d_\lambda}{N!} \Tr[\hat{j}(\lambda)].
\end{align}

\section{Relationships between $j$, $J$ and $\varrho_{\mathrm{int}}$}
\label{app:jJrhoint}

We start by proving the relationship \eqref{eq:jJ} between the functions $j$ and $J$. We begin with Eq.~\eqref{eq:rhotot}:
\begin{align}
	\varrho_{\text{tot}}= \mathcal{N} \hat{S}_\epsilon \left(   \ket{\bm{i}}\bra{\bm{i}} \otimes \varrho_\mathrm{int} \right) \hat{S}_\epsilon,
\end{align}
where
\begin{align}
	\hat{S}_\epsilon= \frac{1}{N!}\sum_{\sigma\in S_N} \epsilon(\sigma) \hat{R}(\sigma) \otimes \hat{R}_\mathrm{int}(\sigma).
\end{align}
Taking the trace over the internal degrees of freedom, we obtain
\begin{align}
	\varrho &=\Tr_{\mathcal{I}^{\otimes N}}(\varrho_{\mathrm{tot}})=\frac{\mathcal{N}}{(N!)^2} \sum_{\sigma,\tau\in S_N} \epsilon(\sigma)\epsilon(\tau) \Tr\left[ \hat{R}_\mathrm{int}(\sigma) \varrho_\text{int} \hat{R}_\mathrm{int}(\tau) \right]  \hat{R}(\sigma) \ket{\bm{i}}\bra{\bm{i}} \hat{R}(\tau),
\end{align}	
such that, for $\pi\in S_N$,
\begin{align}
	j(\pi)&= \frac{N!}{|\stab(\bm{i})|}\braket{\bm{i}|\hat{R}(\pi)^\dagger \varrho|\bm{i}}\notag\\
	&=\frac{\mathcal{N}}{N!|\stab(\bm{i})|} \sum_{\sigma,\tau\in S_N} \epsilon(\sigma)\epsilon(\tau) \Tr\left[ \hat{R}_\mathrm{int}(\sigma) \varrho_\text{int}\hat{R}_\mathrm{int}(\tau) \right]
	\braket{\bm{i}|\hat{R}(\pi)^\dagger\hat{R}(\sigma) |\bm{i}}	\braket{\bm{i}|\hat{R}(\tau) |\bm{i}}\notag\\
	&=\frac{\mathcal{N}|\stab(\bm{i})|}{N!}  \sum_{\sigma,\tau\in S_N}  \epsilon(\tau\circ\sigma) \Tr\left[\varrho_\text{int}\hat{R}_\mathrm{int}(\tau\circ\sigma)  \right] I_{\bm{i}}(\sigma^{-1}\circ \pi)  I_{\bm{i}}(\tau^{-1})\notag\\
	&=\frac{\mathcal{N}|\stab(\bm{i})|}{N!}  \sum_{\sigma,\tau\in S_N}  J(\tau\circ\sigma) I_{\bm{i}}(\sigma^{-1}\circ \pi)  I_{\bm{i}}(\tau^{-1}).
\end{align}
In the last step, we have used the definition \eqref{eq:J} of $J$.
To make convolution products \eqref{convol} appear, we replace $\tau\to\tau^{-1}$ in the sum and obtain
\begin{align}
	j(\pi)=\frac{\mathcal{N}|\stab(\bm{i})|}{N!}  \sum_{\sigma,\tau\in S_N}   I_{\bm{i}}(\tau)J(\tau^{-1}\circ\sigma) I_{\bm{i}}(\sigma^{-1}\circ \pi)=\frac{\mathcal{N}|\stab(\bm{i})|}{N!} (I_{\bm{i}}* J * I_{\bm{i}})(\pi).
\end{align}
To ensure that $\Tr(\varrho_{\mathrm{tot}})=1$, the normalization constant must satisfy
\begin{align}
	\mathcal{N}^{-1}&=\Tr\left[ \hat{S}_\epsilon \left(   \ket{\bm{i}}\bra{\bm{i}} \otimes \varrho_\mathrm{int} \right) \hat{S}_\epsilon   \right]\notag\\
	&=\Tr\left[ \hat{S}_\epsilon \left(   \ket{\bm{i}}\bra{\bm{i}} \otimes \varrho_\mathrm{int} \right)  \right]\notag\\
	&= \frac{1}{N!}\sum_{\sigma\in S_N} \epsilon(\sigma) \braket{\bm{i}|\hat{R}(\sigma)|\bm{i}} \Tr\left[\hat{R}_\mathrm{int}(\sigma)\varrho_\text{int}  \right]\notag\\
	&=\frac{|\stab(\bm{i})|}{N!} (I_{\bm{i}}, J),
\end{align}
such that we indeed have
\begin{align}
	j=\frac{ I_{\bm{i}}* J * I_{\bm{i}}}{( I_{\bm{i}}, J)}.
\end{align}

We now go on to show that any state $\varrho$ supported on the subspace spanned by $\{\hat{R}(\sigma)\ket{\bm{i}}|\sigma\in S_N\}$, and commuting with all permutation operators $\hat{R}(\sigma),\, \sigma\in S_N$,  can be obtained as the reduced external state of a state $\varrho_\mathrm{tot}$ of the form given in Eq.~\eqref{eq:rhotot}.
Actually, rather than $\varrho$, we will recover the function $j$ uniquely associated with it through Eqs.~\eqref{eq:j} and \eqref{eq:rhoj}. For this, we construct an internal state $\varrho_\mathrm{int}$ and compute the corresponding function $J$ with Eq.~\eqref{eq:J}. We then show that $I_{\bm{i}}* J * I_{\bm{i}}/( I_{\bm{i}}, J)$ yields the desired function $j$.
A suitable state $\varrho_\mathrm{int}$ is given by
\begin{align}
	\varrho_\mathrm{int} =\frac{1}{N!}\sum_{\sigma,\tau\in S_N} \epsilon(\sigma\circ\tau^{-1})j(\sigma\circ\tau^{-1}) \hat{R}_\mathrm{int}(\tau)\ket{\bm{s}}\bra{\bm{s}}\hat{R}^\dagger_\mathrm{int}(\sigma),
\end{align}
where $\ket{\bm{s}}=\ket{1,2,\dots N} \in \mathcal{H}_\mathrm{int}^{\otimes N}$ is a tensor product of $N$ mutually orthogonal single-particle internal states (see also the supplemental material of \cite{shchesnovich_universality_2016}).
Indeed, for $\pi\in S_N$,
\begin{align}
	J(\pi)&=\epsilon(\pi) \Tr\left[\varrho_\text{int}\hat{R}_\mathrm{int}(\pi)  \right]\notag\\
	&=\epsilon(\pi)\frac{1}{N!}\sum_{\sigma,\tau\in S_N} \epsilon(\sigma\circ\tau^{-1})j(\sigma\circ\tau^{-1}) \bra{\bm{s}}\hat{R}^\dagger_\mathrm{int}(\sigma)\hat{R}_\mathrm{int}(\pi)  \hat{R}_\mathrm{int}(\tau)\ket{\bm{s}}. 
\end{align}
Given that $\bra{\bm{s}}\hat{R}^\dagger_\mathrm{int}(\sigma)\hat{R}_\mathrm{int}(\pi)  \hat{R}_\mathrm{int}(\tau)\ket{\bm{s}}=\delta_{\sigma,\pi\circ\tau}$, the double sum reduces to 
\begin{align}
	J(\pi)=\epsilon(\pi)\frac{1}{N!}\sum_{\tau\in S_N} \epsilon(\pi)j(\pi)=j(\pi).
\end{align}
We have shown that $J=j$. Since $I_{\bm{i}}* j= j * I_{\bm{i}}=j$ and $( I_{\bm{i}}, j)=1$, we also have $I_{\bm{i}}* J * I_{\bm{i}}/( I_{\bm{i}}, J)=j$, as desired.

\end{document}